\documentclass{ws-ijmpd}
\usepackage[super,compress]{cite}
\begin{document}

\markboth{Kaspi \& Kramer}
{Radio Pulsars:  The Neutron Star Population \& Fundamental Physics}

%
\catchline{}{}{}{}{}
%
\newcommand{\klesssim}{\mathrel{\hbox{\rlap{\hbox{\lower4pt\hbox{$\sim$}}}\hbox{$<$}}}}
\newcommand{\kgtsim}{\mathrel{\hbox{\rlap{\hbox{\lower4pt\hbox{$\sim$}}}\hbox{$>$}}}}
\newcommand{\mpp}{m_{\rm p}}
\newcommand{\mcc}{m_{\rm c}}
\newcommand{\pb}{P_{\rm b}}

\title{Radio Pulsars:  The Neutron Star Population \& Fundamental Physics} 

\author{VICTORIA M. KASPI}

\address{Department of Physics, McGill University, Rutherford Physics Building\\
McGill Space Institute \\
Montreal, Quebec H3W 2C4, Canada \\
vkaspi@physics.mcgill.ca}

\author{MICHAEL KRAMER}

\address{Max-Planck-Institut f\"ur Radioastronomie, Auf dem H\"ugel
  69, 53121 Bonn, Germany\\
Jodrell Bank Centre for Astrophysics, University of Manchester,
Alan-Turing-Building, Manchester M13 9PL, UK\\
mkramer@mpifr.de}

\maketitle

\begin{history}
\received{Day Month Year}
\revised{Day Month Year}
\end{history}


\begin{abstract}
  Radio pulsars are unique laboratories for a wide range of
  physics and astrophysics. Understanding how they are created, how they
  evolve and where we find them in the Galaxy, with or without binary
  companions, is highly constraining of theories of stellar and binary
  evolution. Pulsars' relationship with a recently discovered variety of apparently different
  classes of neutron stars is an interesting modern astrophysical puzzle which we
  consider in Part I of this review.
  Radio pulsars are also famous for allowing us to probe the laws
  of nature at a fundamental level.  They act as precise cosmic clocks and, when
  in a binary system with a companion star,
  provide indispensable venues for precision tests of
  gravity. The different applications of radio pulsars for fundamental physics will be discussed
  in Part II.  We finish by making mention of the newly discovered class of astrophysical
  objects, the Fast Radio Bursts, which may or may not be related to radio pulsars or
  neutron stars, but which were discovered in observations of the latter.
\end{abstract}


\ccode{PACS numbers: 97.60.Gb, 97.60.Jd, 04.80.Cc}


\section{Introduction}

The discovery of evidence for the neutron by Chadwick in 1932
was a major milestone in physics\cite{cha32}, and was surely
discussed with great excitement at the 1933 Solvay Conference titled
``Structure et propri\'et\'es des noyaux atomiques.'' That same year,
two now-famous astronomers, Walter Baade and Fritz Zwicky, suggested
the existence of {\it neutron stars,} which they argued were formed
when a massive star collapses in a ``super-nova'' \cite{bz34b}.
They argued that such a star ``may possess a very small radius and an
extremely high density.'' It took over 3 decades for this seemingly
prophetic prediction to be confirmed: in 1967\footnote{Coincidentally,
the year both of these authors were born.}  then-graduate student Jocelyn Bell
and her PhD supervisor Antony Hewish detected the first radio pulsar
\cite{hbp+68}, and Shklovksy suggested that the X-ray source Sco X-1
was an accreting neutron star \cite{shk67}.  The focus of this paper
is on the former discovery, now a class of celestial objects of which over
2300 are known in our Galaxy (although the accreting variety will be mentioned in
\S\ref{sec:binary}).  Though radio pulsars were compellingly identified
as neutron stars not long after their discovery \cite{gol69,pac68},
the radio emission was unexpected, prompting the
noted physicist and astronomer John Wheeler to remark his surprise 
that neutron stars come equipped with a handle and a bell\footnote{This quote appears
in Ref.~\refcite{mt77} but its origin is unspecified.}.
Though the origin of the radio emission is not well understood
today, it has nevertheless served as a valuable beacon with which we
have learned vast amounts about the neutron star phenomenon. Using
this beacon as a tool also provides us with unique laboratories to
study fundamental physics. In this first part of this contribution, we
will review the diversity in the ``neutron star zoo,'' before we
discuss their applications for understadning the laws of nature, in 
particular gravity, in the second part.

\begin{figure}[b]
\centerline{\psfig{file=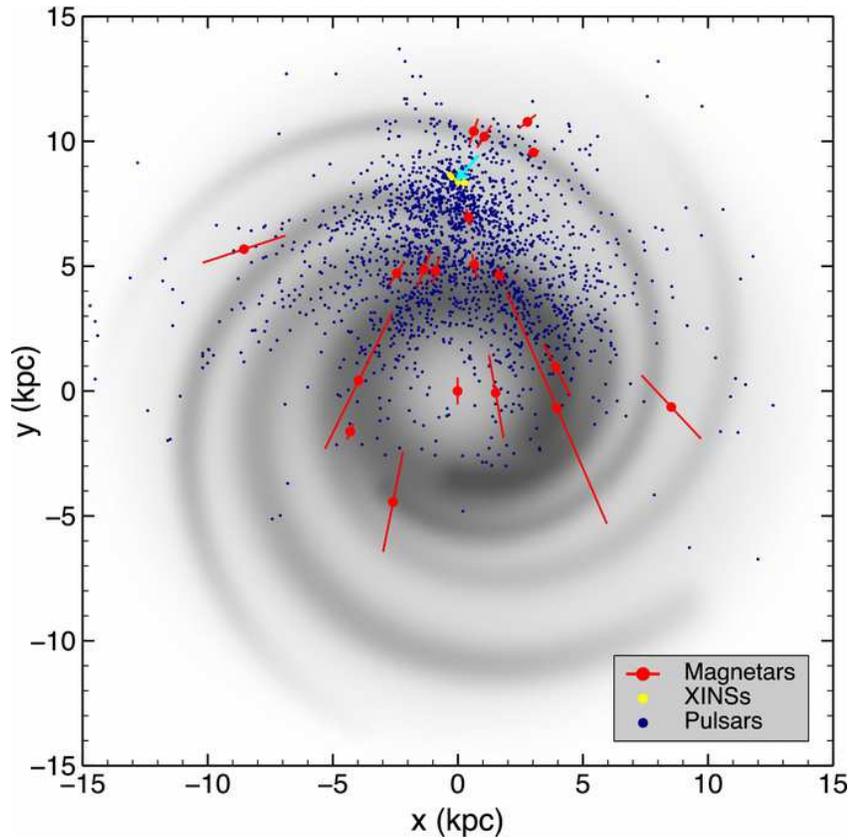,width=11cm}}
\vspace*{8pt}
\caption{Spatial distribution of radio pulsars (in blue), magnetars (in red; see \S\ref{sec:magnetars}), 
XDINS (aka XINS, in yellow; see \S\ref{sec:xdins}),
projected on the Galactic disk. 
The location of the Earth is indicated by a cyan arrow.  The underlying grey scale roughly traces
the free electron distribution.  Figure taken from Olausen \& Kaspi (2014).
\label{fig:nsdist}}
\end{figure}

\section*{Part I: The Different Manifestations of Neutron Stars}

\section{Radio Pulsars}
\label{sec:pulsars}

Radio pulsars are rapidly rotating, highly magnetized neutron stars whose magnetic and rotation axes
are significantly misaligned.  It is believed that beams of radio waves emanate from the magnetic
pole region and are observed as pulsations to fortuitously located observers, with one pulse 
per rotation.  In some cases, two pulses per rotation may be visible if the source's magnetic and
rotation axes are nearly orthogonal.  Pulsations are believed to be produced, and occasionally are
observed, across the full electromagnetic spectrum (see \S\ref{sec:emission}), however the vast majority of known pulsars are
observed exclusively in the radio band.  
The known pulsar population, currently consisting of over 2300 sources with numbers
constantly increasing thanks to ongoing radio pulsar surveys\cite{bck+13,lbr+13,lbh+15}, is largely
confined to the Galactic Plane, with a $e^{-1}$ thickness of $\sim$100~pc \cite{fk06}.  However, the pulsar
scale height appears to increase with source age.  This is presumably because 
pulsars are high velocity objects, with space velocities
typically several hundred km/s \cite{ll94,hp97,hllk05,fk06}.  Such high speeds are likely due 
a birth kick imparted at the time of the supernova, due to a combination of binary disruption (for sources
initially enjoying a binary companion) and asymmetry in the supernova explosion itself.
It is important to note that the known radio pulsar population is very incomplete and subject to strong
observational selection biases; this is clear in Figure~\ref{fig:nsdist} wherein the locations of the
radio pulsars on the Galactic disk are seen to be strongly clustered near Earth.  
These biases include those imposed by dispersion and scattering of radio waves by free electrons
in the interstellar medium, by preferential surveying in the Galactic Plane, as well as by
practical limits on time and frequency resolution in radio pulsar surveys.
See for example
Refs.~\refcite{fk06,blrs14} for a discussion of selection effects in radio pulsar surveys.

Pulsars rotate rapidly, typically with rotation periods $P$ of a few hundred ms.  The presently known
slowest radio pulsar has a period of $\sim$8~s,\cite{ymj99} while the fastest has period 1.4 ms or
716 Hz.\cite{hrs+06}.  All pulsars spin down steadily, a result of magnetic dipole braking, hence can be characterized
by period $P$ and its rate of change $\dot{P}$.  The latter, though typically only tenths of microseconds
per year, is eminently measurable for all known sources because of pulsars' famous rotational stability.
Measurements of pulsar spin-down rate $\dot{P}$ are extremely useful, as they enable helpful 
estimates of key physical properties.  

\begin{figure}[tb]
\centerline{\psfig{file=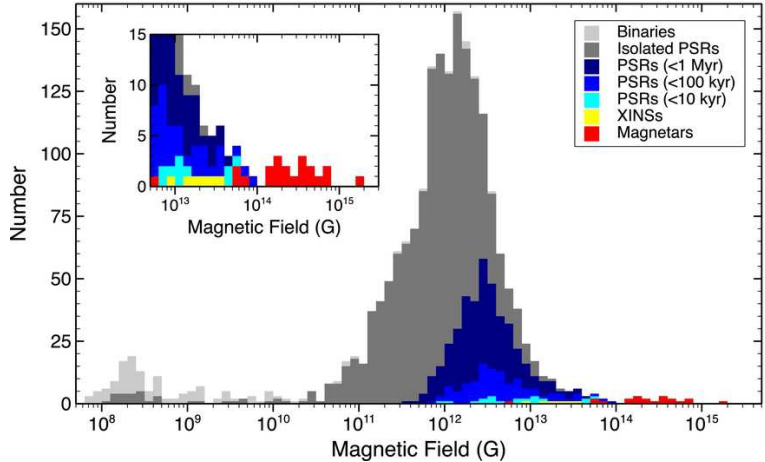,width=10cm}}
\vspace*{8pt}
\caption{Distribution of spin-inferred magnetic fields (using Eq.~\ref{eq:bfield}) of radio pulsars (in various shades of blue and grey), XDINS (aka XINS, in yellow; see \S\ref{sec:xdins}),
and magnetars (in red; see \S\ref{sec:magnetars}). From Olausen \& Kaspi (2014).
\label{fig:bdist}}
\end{figure}

One such property is the surface dipolar magnetic field at the equator,
\begin{equation}
B = \left( \frac{3c^3I}{8 \pi^2 R^6} \right)^{1/2} \sqrt{P \dot{P}} = 3.2 \times 10^{19} \;\; \sqrt{P \dot{P}} \;\;\;  G,
    \label{eq:bfield}
\end{equation}
where $I$ is the stellar moment of inertia, typically estimated to be
$10^{45}$~g~cm$^2$, and $R$ is the neutron star radius, usually
assumed to be 10 km (see Part II. for observational constraints.) This estimate assumes
magnetic braking {\it in vacuo}, which was shown to be impossible early in the history of these objects \cite{gj69}
since rotation-induced electric fields dominate over the gravitational force, even for these compact stars,
such that charges must surely be ripped from the stellar surface and form a dense magnetospheric plasma.  Nevertheless,
modern relativistic magnetohydrodynamic simulations of pulsar magnetospheres have shown that the
simple estimate offered by Eq.~\ref{eq:bfield} is generally only a factor of 2--3 off \cite{spi06}.
The observed distribution of radio pulsar magnetic fields is shown in Figure~\ref{fig:bdist}.

Measurement of $P$ and $\dot{P}$ also enable an estimate of the source's age.
The pulsar's characteristic age $\tau_c$ is defined as
\begin{equation}
\tau_c = \frac{P}{(n-1)\dot{P}} \left[ 1- \left( \frac{P}{P_0}  \right)^{(n-1)} \right] \simeq \frac {P}{2\dot{P}},
\label{eq:tauc}
\end{equation}
where $n$ is referred to as the `braking index' and is equal to 3 for simple
magnetic dipole braking (see e.g. Ref.~\refcite{mt77}), though is observed to be less than 3 in
the handful of sources for which a measurement of $n$ has been possible.\cite{lkg+07}
$P_0$ is the spin period at birth and is generally assumed to be much smaller than the 
current spin period, although this is not always a valid assumption, particularly for
young pulsars \cite{krv+01}.  

Finally, a pulsar's spin-down luminosity
$L_{sd}$ (also known as $\dot{E}$, where $E \equiv \frac{1}{2} I \omega^2$ with $\omega \equiv 2\pi/P$
is the stellar rotational kinetic energy) can be estimated from $P$ and $\dot{P}$ and is given by
\begin{equation}
\dot{L}_{sd} = \frac{d}{dt} \left( \frac{1}{2} I \omega^2 \right) = I \omega \dot{\omega} = 4 \pi^2 I \frac{\dot{P}}{P^3} = 4 \times 10^{31} \left( \frac{\dot{P_{-15}}}{P_1} \right) \;\;\; {\rm erg/s},
    \label{eq:edot}
\end{equation}
where $\dot{P}_{-15}$ is $\dot{P}$ in units of $10^{-15}$ and $P_1$ is the period in units of seconds.
$L_{sd}$ represents the power available for conversion into electromagnetic radiation, an upper limit on
the (non-thermal; see \S\ref{sec:emission}) radiation a pulsar can produce.  For this reason, radio pulsars
are also known as `rotation-powered pulsars.'

\begin{figure}[tb]
\centerline{\psfig{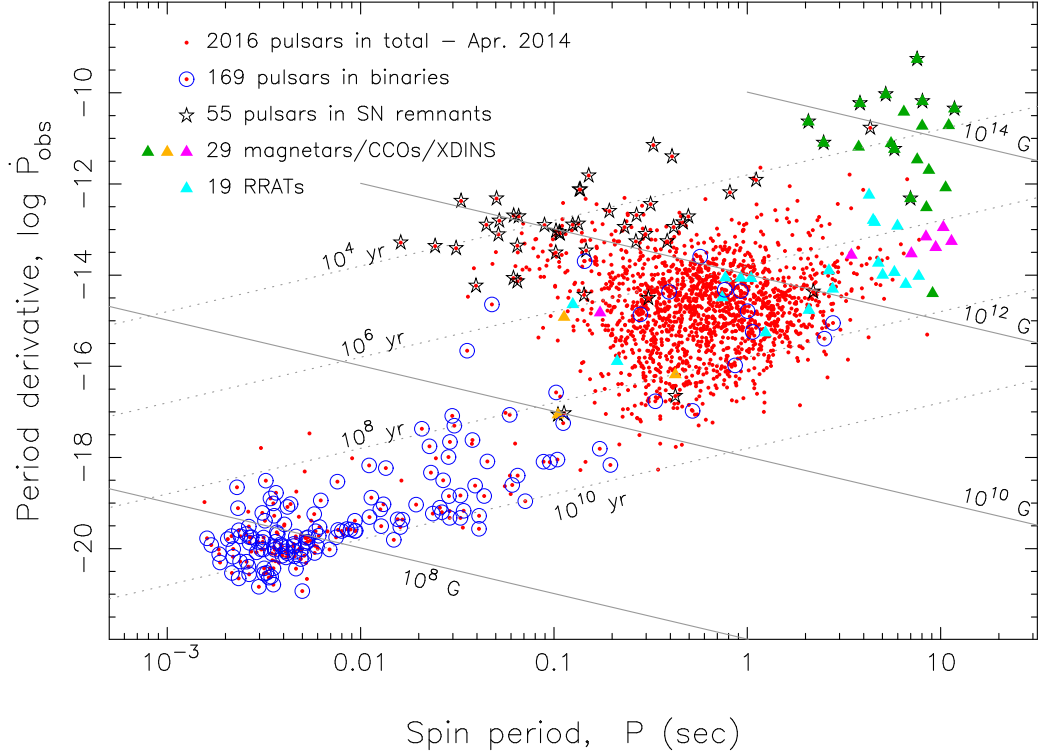}}
\vspace*{8pt}
\caption{$P$-$\dot{P}$ diagram.  Red dots indicate known radio pulsars as of April 2014.
    Blue circles represent binaries.  Stars represent associated supernova remnants.  Magnetars are
    represented by green triangles (see \S\ref{sec:magnetars}).  The XDINS and CCOs (see \S\ref{sec:diversity},\ref{sec:magnetars})
    are pink and yellow triangles, respectively.  RRATs (\S\ref{sec:rrats}) are in cyan. Solid grey lines are of constant
    magnetic field (Eq.~\ref{eq:bfield}) and dotted lines are of constant characteristic age (Eq.~\ref{eq:tauc}).
    From Tauris et al. (2014).
\label{fig:ppdot}}
\end{figure}

A traditional way of summarizing the pulsar population is via the $P$-$\dot{P}$ diagram (Fig.~\ref{fig:ppdot}).
Here the spin periods of pulsars are plotted on the $x$-axis and $\dot{P}$ on the $y$-axis.  The swarm of conventional radio
pulsars clearly has its $P$ peak near $\sim$500 ms, with typical $B \simeq 10^{11}$~G and characteristic age
$\tau_c \simeq 10^7$~yr.  The youngest radio pulsars have $\tau_c \simeq 1$~kyr and are generally found in
supernova remnants; the latter dissipate typically after 10--100~kyr, explaining why older pulsars are generally
not so housed in spite of all having been born in core-collapse supernovae.  The $P=33$-ms Crab pulsar is arguably
the most famous of the young pulsars, its birth having been recorded by Asian astrologers in 1054 A.D.\cite{cs77a} 
However, it is in fact not the youngest known pulsar; this honour presently goes to the 884-yr old PSR J1846$-$0258 in the supernova remnant Kes 75.\cite{lkgk06}
Also clear in the $P$-$\dot{P}$ diagram is the collection of binary pulsars, nearly all of which cluster in the lower
left portion of the diagram, where the ``millisecond pulsars'' reside.  This is no coincidence; although
the rapid rotatation of the young Crab-like pulsars is almost certainly a result of angular momentum conservation
in the core collapse, that of the millisecond pulsars (MSPs) is intimately tied to their binarity.  MSPs are believed
to have been spun-up by an episode of mass accretion from their binary companion (see \S\ref{sec:binary}).

\subsection{Pulsar Emission}
\label{sec:emission}

Though rotation-powered pulsars are usually referred to as `radio pulsars,' in reality these objects
emit across the full electromagnetic spectrum.  In fact, the radio emission (that in the $\sim$100~MHz to $\sim$100~GHz
range), which must surely be
of a non-thermal nature owing to the enormous brightness temperatures implied, usually represents a tiny
fraction (typically $\sim 10^{-6}$) of $L_{sd}$, hence is energetically unimportant.  The
richness of radio observations and phenomenology has fuelled over the years significant theoretical
effort into understanding its origin.  However at present, there is no concensus and it remains an open question
\cite{mel93,mel95}. In spite of the lack
of an understanding of the physics of the radio emission, pulsar astronomers are generally content
to accept its existence as coming from a `black box,' and use it as an incredibly useful beacon of the
dynamical behaviour of the star as described in Part II.

The second most commonly observed emission from rotation-powered pulsars is in the X-ray band.  
See Ref.~\refcite{bec00} or \refcite{krh06} reviews.  The origin of pulsar
X-rays is far better understood than is the radio emission and we describe it briefly here as it is
instructive, particularly when considering other classes of neutron stars (see \S\ref{sec:diversity}).  X-rays originate
from one of two possible mechanisms, which can sometimes both be operating.  One is thermal emission from the surface,
due either to the star being initially hot following its formation in a core-collapse event (in which case
the thermal luminosity need not be constrained by $L_{sd}$), or from surface reheating by return currents
in the magnetosphere.  The latter is particularly common in millisecond pulsars, but may well be present
in all pulsars and indeed can be an important complicating factor in efforts to constrain neutron-star
core composition via studies of cooling.  As the thermal emission is thought to arise from the surface, it
is typically characterized by quasi-sinusoidal pulsations, likely broadened by general relativistic light bending.
The second source of X-rays is purely magnetospheric.  This emission has a strongly non-thermal spectrum and
is appears highly beamed, as observed via very short duty-cycle pulsations.  The non-thermal emission is
ultimately powered by spin-down (as is the thermal return-current emission) so its luminosity must
be limited by $L_{sd}$.  Note that additional X-ray emission can be present in pulsars' immediate vicinity due to
{\it pulsar wind nebulae} -- sometimes spectacular synchrotron nebuale that result from pulsars' relativistic
winds being confined by the ambient medium.  See Ref.~\refcite{gs06} for a review of these objects.

Space limitations preclude discussion of the third-most commonly observed emission band for rotation-powered pulsars --
the gamma-ray regime.  For a recent review of this interesting and highly relevant area of radio pulsar astrophysics,
see Ref.~\refcite{car14}.

\section{Binary Radio Pulsars}
\label{sec:binary}

\begin{figure}[tb]
\centerline{\psfig{file=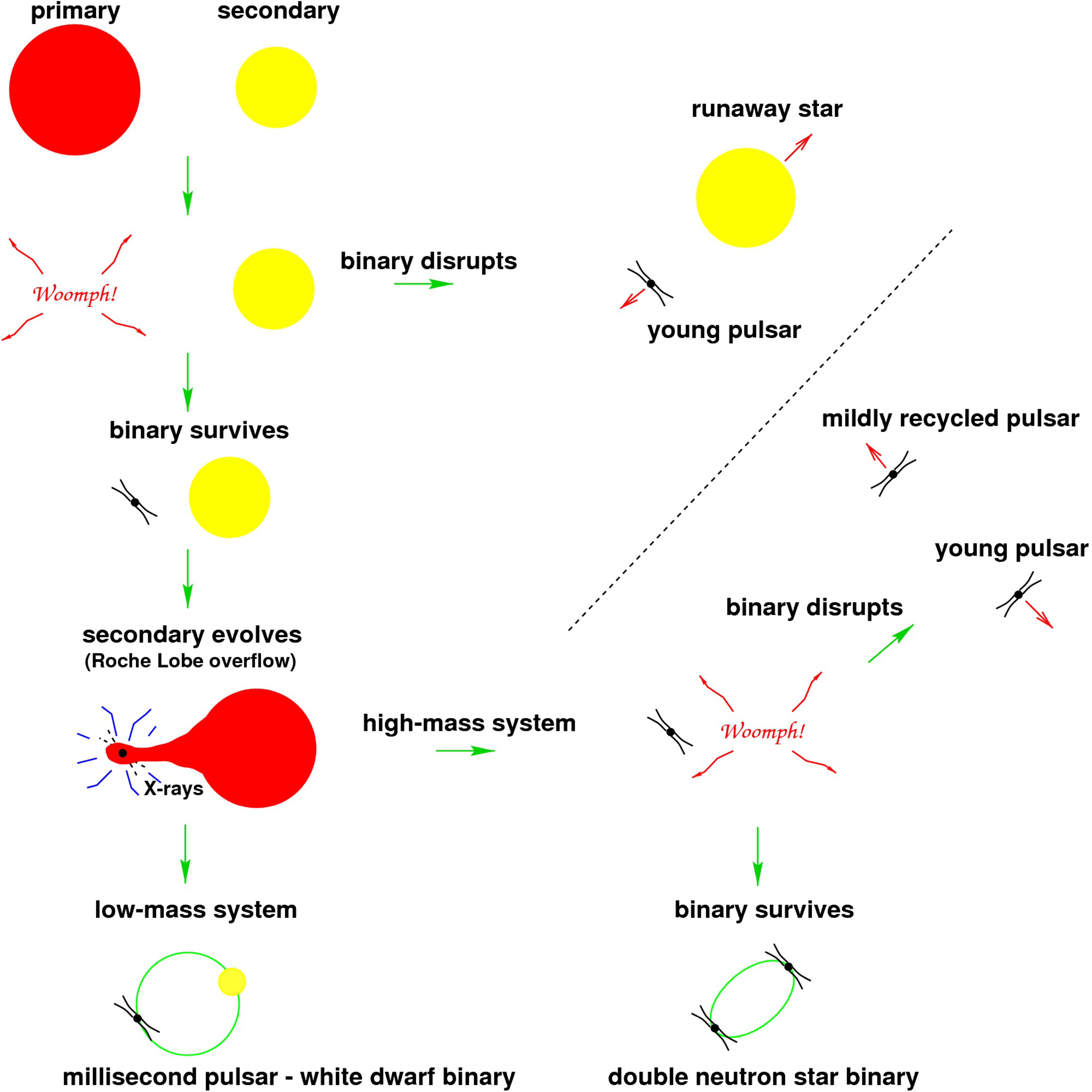,width=9cm}}
\vspace*{8pt}
\caption{Two neutron star binary evolution scenarios, one forming a millisecond pulsar -- white dwarf binary,
and the other a double neutron star binary.  The primary deciding factor in the end state of the neutron
star is the mass of its companion, with the white dwarf binary forming from a low-mass companion 
and the double neutron star from a high-mass companion.  From Lorimer (2008).
\label{fig:evolution}}
\end{figure}

As seen in Figure~\ref{fig:ppdot}, pulsars with a binary companion generally, but not exclusively, inhabit the lower
left of the $P$-$\dot{P}$ diagram, where spin periods are short and spin-down rates low.  Indeed the vast
majority of millisecond pulsars are in binary systems and have among the lowest magnetic field strengths
of the pulsar population (see the small peak at the very low end in the $B$-field distribution in Fig.~\ref{fig:bdist}).
These facts are not coincidental.  According to the standard model\cite{bv91,pk94,tv06}, 
although the vast majority of radio pulsars originated
from progenitors that were in binaries, most of these systems were disruped by the supernova.  Of the few
that survived, subsequent evolution of the pulsar binary companion, under the right circumstances, resulted
in Roche-lobe overflow and the transferring of matter and angular momentum onto the neutron star, in the 
process spinning it up.  Such spun-up pulsars are often called ``recycled'' as they are effectively given a new
life by their companion; without the latter they would have spun down slowly, alone, until ultimately the
radio emission mechanism ceased as it evenutally must.
The mass transfer phase, observed as a bright accreting X-ray source powered
by the release of gravitational energy as the transferred matter falls onto the neutron star\cite{tv06}, 
has a final result that depends strongly
on the nature of the companion and its proximity to the neutron star.  For low-mass companions, this mass transfer
phase can last long enough to spin the pulsar up to millisecond periods.  For higher-mass companions, only
tens of millisecond periods can be achieved as these companions have shorter lifetimes.  Simultaneous with the 
spin-up is an apparent quenching of
the magnetic field, a process whose physics are poorly understood, but for which there is strong observational
evidence.  The above is a very broad-brush description of a very rich field of quantitative research that
blends orbital dynamics with stellar evolution and neutron-star physics.
One pictoral example of evolutionary scenarios that can lead to the formation of recycled pulsars is shown
in Figure~\ref{fig:evolution}.\cite{lor08}

One outstanding mystery in the standard evolutionary model is the existence of isolated MSPs.
These can be seen scattered in the lower left-hand part of Figure~\ref{fig:ppdot}.  
Indeed, the first discovered MSP, PSR B1937+21, is isolated.\cite{bkh+82}
If binarity is key to recycling and spin-up, where are the isolated MSPs companions?
One plausible answer may lie in the apparent companion `ablation' that appears to be in progress
in some close (orbital periods of a few hours) MSP binaries, notably those in which the 
radio pulsar is regularly eclipsed by material that extends well beyond the surface
of the companion\cite{fst88,sbl+96,asr+09}.  The companion's mass loss is believed to be
fueled by the impingement of the intercepted relativistic pulsar particle wind which is ultimately
powered by $L_{sd}$. 

Another newly identified mystery is the discovery of eccentric MSP binaries. 
Key to the recycling process is
rapid and efficient circularization of orbits and indeed some MSP binaries\cite{lfl+06,vbc+09} 
have eccentricities well under $10^{-6}$.  The discovery of a eccentricity 0.44 MSP in a 95-day orbit
in the Galactic disk\cite{crl+08} thus was difficult to understand; one possibility is that it formed as part of
a hierarchical triple system\cite{fbw+11} in which the inner companion was eventually ablated.  The
recent unambiguous detection of an MSP in a hierarchical triple system\cite{rsa+14} supports the existence
of such systems, and suggests that binary evolution may be an incomplete picture of the paths to making
MSPs\cite{tv14}.  Today there are 3 more MSP binary systems having eccentricities $\kgtsim$0.1 (Refs.
\refcite{dsm+13,bck+13,kls+15}) and though
origins in triple systems are still on the table, 
other scenarios for their production, including accretion-induced collapse of a super-Chandrasekhar
mass oxygen-neon-magnesium white dwarf in a close binary \cite{ft14} 
and dynamical interaction with a circumbinary disk \cite{ant14}, have been proposed.

Very recently, there has been a series of spectacular confirmations of key aspects of binary evolution theory.
One is
in the form of the discovery of a binary radio MSP, PSR J1023+0038, which had been
observed to have an accretion disk in the previous decade\cite{asr+09}.  Then there came the discovery
of repeated swings between radio pulsations and bright accretion-powered X-ray pulsations
in a different source\cite{pfb+13}.  Interestingly, the radio pulsations from 
PSR J1023+0038 have subsequently vanished\cite{sah+14} and a far brighter X-ray source
has turned on\cite{tya+14}, suggesting some form of accretion, possibly in the propeller
regime, is ongoing.  Yet a third similar X-ray binary/radio MSP transitioning source has also recently been
identified\cite{bpa+14}.  This flurry of discoveries has brought us into a new era for making progress on
the physics of accretion and accretion flows, the nature of the end of the recycling process
and the formation of radio MSPs.

Finally, it is important to note the handful of radio pulsar binaries that sit among the regular
population in the $P$-$\dot{P}$ diagram, i.e. young binaries in which the pulsar has not yet been
recycled, and in which the companion is a massive main-sequence star.  
Only a few such objects are known\cite{jml+92,kjb+94,sml+01} likely owing to their short lifetimes.
Unsurprisingly, these binaries are highly eccentric, resulting from a kick likely imparted at the
time of the supernova explosion that formed the pulsar, but which (barely) did not unbind the orbit.
These systems are interesting for a variety of reasons, including unusual dynamics present due to 
spin-induced quadrupole moments in the massive star, such as coupling between the stellar and orbital
angular momenta.\cite{lbk95}  This can cause precession in the system which can be used to detect misalignment
between the stellar and orbital angular momenta, which provides strong evidence for a kick at the
time of the neutron-star formation.\cite{kbm+96,wex98,msk+12}   Also, these systems provide a unique way to
constrain the nature of massive star winds\cite{jbwm05,ktm96}.  One is also a $\gamma$-ray
emitter\cite{aaa+11,haa+13}, and serves as a possible `Rosetta Stone' for a different class of $\gamma$-ray-emitting
binaries in which the nature of the compact object is unknown.\cite{mir12}

\section{Diversity in Neutron Stars}
\label{sec:diversity}

The last decade has shown us that the observational properties of
neutron stars are remarkably diverse:  Wheeler's `handle and bell,'
invoked to describe emission from radio pulsars, now appears to be
occasionally accompanied or sometimes substituted by a horn, a basket,
a flashing light and/or a flag.  It turns out, radio pulsars are
just one observational manifestation of neutron stars.  Today we have
identified multiple other classes (or possibly sub-classes): magnetars
(which have been sub-classified into `anomalous X-ray pulsars (AXPs)'
and `soft gamma repeaters (SGRs)'), X-ray dim isolated neutron stars
(XDINS), Central Compact Objects (CCOs), and Rotating Radio Transients
(RRATs).  In addition to an explosion of acronyms, we have an explosion
of phenomenology. See Ref.~\cite{kas10} for a review.
An important current challenge in neutron-star
astrophysics is to establish an overarching physical theory of neutron
stars and their birth properties that can explain this great diversity.
Next we discuss each of these classes in turn.

\section{Magnetars}
\label{sec:magnetars}

Magnetars are without doubt the most dramatic of the neutron star population,
with their hallmark observational trademark the emission of brief but intense -- often
greatly hyper-Eddington -- X-ray and soft $\gamma$-ray bursts.  
This class of neutron stars was first noted in 1979 with the detection of repeated
soft $\gamma$-ray bursts from two different sources by space-based detectors \cite{mgg79,mgi+79} -- hence the name
`soft gamma repeater' (SGR).  Today there are 23 confirmed magnetars; 
the first magnetar catalog has been published \cite{ok14} and
is available online\footnote{http://www.physics.mcgill.ca/$\sim$pulsar/magnetar/main.html}.
See Ref.~\refcite{mpm15} for a very recent review.
Three magnetars have shown particularly powerful ``giant flares;''\cite{hbs+05,pbg+05} in the first 0.2 s of
one such event, from SGR 1806$-$20, 
more energy was released than the Sun produces in a quarter of a million years \cite{hbs+05}
and in the first 0.125 s, the source outshone by a factor of 1000 all the stars in
our Galaxy, with peak luminosity upwards of $2\times 10^{47}$~erg~s$^{-1}$ and total
energy released approximately $4\times 10^{46}$ erg.

Apart from their signature X-ray and soft $\gamma$-ray bursts, magnetars have
the following basic properties.  They are persistent X-ray pulsars, with periods for known
objects in the range 2--12~s and are all spinning down, such that application of the
standard magnetic braking formula (Eq.~\ref{eq:bfield}) yields field strengths typically
in the range $10^{14}$-$10^{15}$~G.  In the past, two sub-classes have been referred to
in the literature:  the SGRs, and the `anomalous X-ray pulsars' (AXPs) which, prior to 2002,
had similar properties to the SGRs except did not seem to burst (but see below).
Roughly 1/3 of all these sources are in supernova remnants, which clearly
indicates youth; in very strong support of this is the tight confinement of Galactic
magnetars (two are known in the Magellanic Clouds) to the Galactic Plane, with a scale
height of just 20--30~pc \cite{ok14}.  This, along with some magnetar associations with
massive star clusters\cite{mcc+06}, strongly suggests that magnetars are preferentially
produced by very massive ($\kgtsim$30M$_{\odot}$) stars that might otherwise have naively
have been though to produce black holes.  Note that
the magnetar spatial distribution in the Galaxy is subject to far fewer selection effects than
is that of radio pulsars (see Fig.~\ref{fig:nsdist}), 
because magnetars are typically found via their hard X-ray bursts (on which the interstellar
medium has no effect) using all-sky
monitors that have little to no preference for direction.

Importantly, and at the origin of their name, is
that in many cases their X-ray luminosities and/or their burst energy outputs (and certainly
the giant flare energy outputs!)
are orders of magnitude larger than what is available from their rotational kinetic energy
loss, in stark contrast with conventional radio pulsars.  Thus the main puzzle regarding
these sources initially was their energy source.  Accretion from a binary companion was ruled out early
on given the absence of any evidence for binarity.\cite{mis98}
Thompson and Duncan \cite{dt92,td95,td96a} first developed the magnetar
theory by arguing that an enormous internal magnetic field would be unstable to decay and
could heat the stellar interior\cite{gr92}, thereby stressing the crust from within, resulting in
occasional sudden surface and magnetospheric magnetic restructuring that could explain
the bursts.  That same high field, they proposed, could explain magnetars' relatively long spin periods in spite
of their great youth, as well as confine the energy seen in relatively long-lived tails of giant 
flares.  The direct measurement of the expected spin-down rates\cite{kds+98} (and the implied
spin-inferred magnetic fields mentioned above) came, crucially, after this key prediction.
This provided the most powerful confirmation of the magnetar model; additional strong
evidence came from the detection of magnetar-like bursts from the AXP source class \cite{gkw02,kgw+03} which
had previously been explicitly called out in Ref.~\refcite{td96a} as being likely magnetars.

Although the magnetar model is broadly accepted by the astrophysical community, as for radio pulsars, a
detailed understanding of their observational phenomena is still under development.  Following the seminal
theoretical work in Refs.~\refcite{dt92,td95,td96a}, later studies have shown that magnetar magnetospheres likely suffer various
degrees of `twisting,' either on a global scale \cite{tlk02} or, more likely, in localised regions that
have come to be called `j-bundles' \cite{bt07}.  The origin of sudden X-ray flux enhancements at the times of outburst
may be in the development of these twists, with subsequent radiative relaxation coupled with field untwisting
\cite{}.  On the other hand, interior heat depositions can also account for the observed flux relaxations
post outburst, and, in this interpretation, can potentially yield information on crustal composition.\cite{let02,kew+03,skc14}.
Interesting open questions surround magnetar spectra, which are very
soft below 10 keV, consisting of a thermal component that is rather hot ($kT \simeq 0.4$~keV) compared
with those of radio pulsars (\S\ref{sec:emission}), and a non-thermal component that may arise from
resonant Compton scattering of thermal photons by charges in magnetospheric currents \cite{}.  A sharp upturn
in the spectra of magnetars above $\sim$15~keV \cite{khm04,khdc06} was unexpected but may be explainable
of coronal outflow of $e^{\pm}$ pairs which undergo resonant scattering with soft X-ray photons and
lose their kinetic energy at high altitude \cite{bel13}.  Another magnetar mystery
is that they are prolific glitchers \cite{dkg08}, in spite of apparently high interior temperatures
that previously were invoked in the young and presumably hot Crab pulsar to explain its paucity
of glitches \cite{ai75,acp89}.  Also, some magnetar glitch properties are qualitatively different from those
of radio pulsars, starting with their frequent (but not exclusive) association with bright X-ray outbursts
\cite{kgw+03,dk14,akn+13}.

\subsection{High-B Radio Pulsars and Magnetars}

\label{sec:1745}
One particularly interesting issue is how especially high-$B$ radio pulsars relate to
magnetars.  Figure~\ref{fig:bdist} shows histograms of the spin-inferred magnetic
field strengths of radio pulsars (coloured by age) and magnetars.  Although, generally speaking,
magnetar field strengths are far higher than those of radio pulsars, there is a small overlap
region in which there exist otherwise ordinary radio pulsars having magnetar-strength fields,
and magnetars having rather low $B$ fields \cite{ret+10,skc14}.  
This is also easy to see in Figure~\ref{fig:ppdot}.  A partial answer to this comes from an event
in 2006 in which the otherwise ordinary (though curiously radio quiet) rotation-powered pulsar PSR J1846$-$0258, albeit
one with a moderately high $B$ of $5 \times 10^{13}$~G, suddenly
underwent an apparent `magnetar metamorphosis,' brightening by a factor of $>20$ in the X-ray band
and emitting several magnetar-like bursts \cite{ggg+08}.  This outburst lasted $\sim$6 weeks, and then
the pulsar returned to (nearly) its pre-outburst state. (See Ref.~\refcite{lnk+11} for the post-outburst status.)
This suggests that in high-$B$ rotation-powered pulsars, there is the capacity for magnetar-type instabilities.
Recent theoretical work on magnetothermal evolution in neutron stars supports this.\cite{pp11,vrp+13}
Conversely, radio emission has now been detected 
from 4 magnetars\cite{crh+06,crhr07,lbb+10,efk+13}, although it has interestingly different properties from that typical of radio pulsars.
Notably it is often more variable, has an extremely flat radio
spectrum, is essentially 100\% linearly polarized and appears to be present only
after outbursts, fading away slowly on time scales of months to years.
One particularly interesting radio magnetar is SGR J1745$-$2900, found in the Galactic Centre, within
$3''$ of Sgr A* \cite{mgz+13,rep+13,efk+13,sj13}.  Though plausibly gravitationally bound to the black hole, its rotational
instabilities (typical for magnetars) will likely preclude dynamical experiments
\cite{efk+13,kab+14}.  Nevertheless it is of considerable interest as its radio emission suffers far less interstellar
scattering than
expected given its location, suggesting future searches of the Galactic Centre region for more rotationally
stable radio pulsars may bear fruit and allow sensitive dynamical
experiments as described in Part II.

\section{XDINS}
\label{sec:xdins}

The `X-ray Dim Isolated Neutron Stars' (XDINS; also sometimes known more simply
as Isolated Neutron Stars, INSs) are sub-optimally named
neutron stars because (i) the term `dim' is highly detector specific,
and (ii) most radio pulsars are both neutron stars and `isolated.'  Nevertheless this name
has stuck and refers to a small class that has the following
defining properties:  quasi-thermal X-ray emission with relatively low X-ray luminosity, great
proximity, lack of radio counterpart, and relatively long periodicities ($P=$3--11~s).
For past reviews of XDINSs, see Refs.~\refcite{krh06,hab07,kv09}.
XDINSs may represent an interestingly large fraction of all Galactic neutron stars \cite{kk08}; we
are presently only sensitive to the very nearest such objects (see Fig.~\ref{fig:nsdist}).
Timing observations of several objects have revealed that they are
spinning down regularly, with inferred dipolar surface magnetic fields of typically a
$\sim 1-3 \times 10^{13}$~G \cite{kv05a,kv09}, and characteristic ages of
$\sim$1--4 Myr (see Fig.~\ref{fig:ppdot}).  Such fields are somewhat higher than the typical
radio pulsar field.  This raises the interesting question of why the closest neutron stars should have
preferentially higher $B$-fields.
The favoured explanation for XDINS properties is that they are
actually radio pulsars viewed well off from the radio beam.  Their X-ray luminosities are thought to
be from initial cooling and they are much less luminous than younger thermally cooling radio pulsars because
of their much larger ages.  However, their luminosities are too large for conventional
cooling, which suggests an addition source of heating, such as magnetic field decay, which
is consistent with their relatively high magnetic fields.

\section{`Grand Unification' of Radio Pulsars, Magnetars and XDINS: Magnetothermal Evolution}
\label{sec:unification}

Recent theoretical work suggests that radio pulsars, magnetars and XDINS can be understood under
a single physical umbrella as having such disparate properties simply because of their
different birth magnetic fields and their present ages.
Motivated largely by mild correlations between spin-inferred $B$ and
surface temperature in a wide range
of neutron stars, including radio pulsars, XDINSs and magnetars
\cite{plmg07}
(but see Ref.~\refcite{zkgl09}), 
a model of `magneto-thermal evolution' in neutron stars has been developed in which
thermal evolution and magnetic field decay are
inseparable \cite{plmg07,apm08b,pmg09,vrp+13,gc14}.
Temperature affects crustal electrical resistivity, which
in turn affects magnetic field evolution, while the decay of the field
can produce heat that then affects the temperature evolution.
In this model, neutron stars born with large magnetic fields ($>5\times 10^{13}$~G)
show significant field decay, which keeps them hotter longer.  The magnetars
are the highest $B$ sources in this picture, consistent with observationally
inferred fields;  the puzzling fact that XDINSs, in spite of their
great proximity, appear to have high inferred $B$s relative to radio pulsars
is explained as the highest $B$ sources
remain hottest, hence most easily detected, longest.  


\section{CCOs}
\label{sec:ccos}

\begin{figure}[bt]
\centerline{\psfig{file=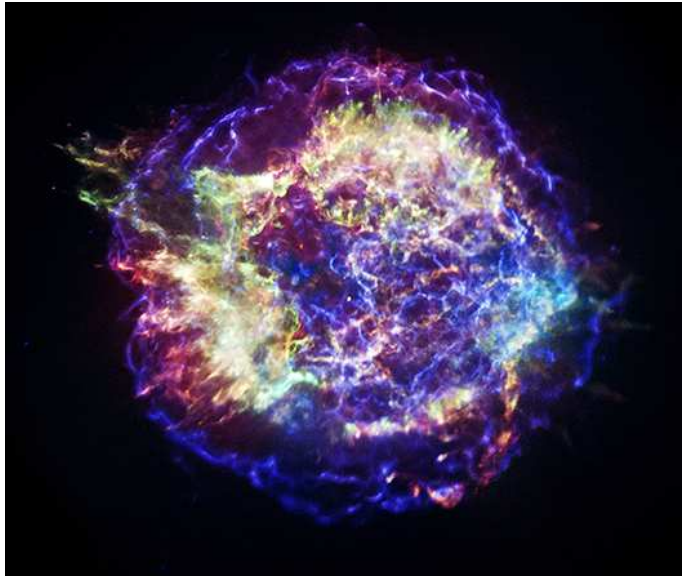,width=9cm}}
\vspace*{8pt}
\caption{X-ray image of the Cas A supernova remnant, obtained with the {\it Chandra} X-ray
Observatory, showing the mysterious compact object at the centre.  Image from
http://chandra.harvard.edu/photo/2013/casa/.
\label{fig:casa}}
\end{figure}

A census of neutron-star classes should mention the so-called Central
Compact Objects (CCO)\footnote{Again, a rather poor name that has stuck:  the Crab pulsar is certainly `central' to its nebula and compact, nevertheless is {\it not} considered a CCO!}.
CCOs are a small, heterogeneous collection of 
X-ray emitting neutron-star-like objects at the centres of supernova remnants,
but having puzzling properties which defy a clean classification.  Properties common
among CCOs are absence both of associated nebulae and of counterparts at other
wavelengths.
The poster-child CCO, discovered in the first-light observation of the {\it Chandra} observatory (Fig.~\ref{fig:casa}),
is the mysterious central object in the young oxygen-rich
supernova remnant Cas A.  Particularly
puzzling is its lack of X-ray periodicity, lack of associated nebulosity,
and unusual X-ray spectrum
\cite{pza+00,cph+01,mti02,pl09}.  

Other objects that have been previously designated CCOs have been revealed to have low-level
X-ray pulsations and surprisingly small spin-down rates.
PSR J1852+0040 is at the centre of
the SNR Kes 79 \cite{ghs05,hgcs07}.  This undoubtedly young pulsar, observed only in
X-rays, has $P=105$~ms yet a magnetic
field strength of only $B = 3.1 \times 10^{10}$~G.  Its characteristic age,
$\tau_c=192$~Myr \cite{hg10}, is many orders of magnitude larger
than the SNR age, and much older than would be expected for an object
of its X-ray luminosity (which greatly exceeds the spin-down luminosity).
Interestingly, the object sits in a sparsely populated region of the $P$-$\dot{P}$ diagram
(Fig.~\ref{fig:ppdot}), among mostly recycled binary pulsars.
A similar case is the CCO in the SNR
PKS 1209$-$52, 1E 1207.4$-$5209.  This 0.4-s X-ray pulsar 
\cite{zpst00,pzst02} has a spin-down rate that implies
$B=9.8 \times 10^{10}$~G and age again orders of magnitude greater
than the SNR age and inconsistent with a so large X-ray luminosity \cite{gha13}.
Yet another such low-$B$ CCO is RX J0822$-$4300 in Puppis A \cite{gh09},
with $P=112$~ms and $B=2.9 \times 10^{10}$~G \cite{gha13}.
Ref.~\refcite{hg10} presents a synopsis of other sources classified as CCOs and argues that they are X-ray bright thanks to residual thermal cooling
following formation, with the neutron star having been born spinning
slowly.  If so, the origin of the
non-uniformity of the surface thermal emission is hard to understand.
Even more puzzling however is the very high implied birthrate of these
low-$B$ neutron stars coupled with their extremely slow spin-downs:  although
none of these objects has yet shown radio emission, if one did, it should `live'
a very long time compared to higher-$B$ radio pulsars, yet the region of the
$P$-$\dot{P}$ diagram where CCOs should evolve is greatly underpopulated in spite
of an absence of selection effects against finding them (see also Ref.~\refcite{kas10}).
This argues that for some reason, CCO-type objects must never become radio pulsars,
which is puzzling, as there exist otherwise ordinary radio pulsars with 
CCO-like spin properties.

\section{Rotating Radio Transients}
\label{sec:rrats}

No neutron-star census today is complete without a discussion of the so-called Rotating Radio Transients,
or RRATs.  RRATs are a curious class of Galactic radio sources \cite{mll+06}
in which only occasional pulses are detectable, with conventional periodicity searches showing
no obvious signal.  Nevertheless, the observed pulses are inferred to occur at multiples of an underlying periodicity
that is very radio-pulsar-like.  Indeed, patient RRAT monitoring has shown that they also spin down at rates similar
to radio pulsars.
The number of known RRATs is now approximately 90\footnote{See the online ``RRatalog'' at http://astro.phys.wvu.edu/rratalog/rratalog.txt}.
although just under 20 have spin-down rates measured.  At first thought
to be possibly a truly new class of neutron star, it now appears most
reasonable that RRATs are just an extreme form of radio pulsar, which have long
been recognized as exhibiting sometimes very strong modulation of their
radio pulses \cite{wsrw06,kkl+11}.  Indeed several RRATs sit in unremarkable
regions of the $P$-$\dot{P}$ diagram (Fig.~\ref{fig:ppdot}).
Interesting though is the mild evidence for
longer-than-average periods and higher-than-typical $B$ fields among the RRATs than in the
general population. 
Regardless of whether RRATs are substantially physically different
from radio pulsars, their discovery is important as it suggests a large population
of neutron stars that was previously missed by radio surveys which looked only
for periodicities.  This may have important implications for the neutron-star
birth rate and its consistency with the core-collapse supernova rate \cite{mll+06,kk08}.


\section{Fast Radio Bursts: A New Mystery}
\label{sec:frbs}

\begin{figure}[t]
\centerline{\psfig{file=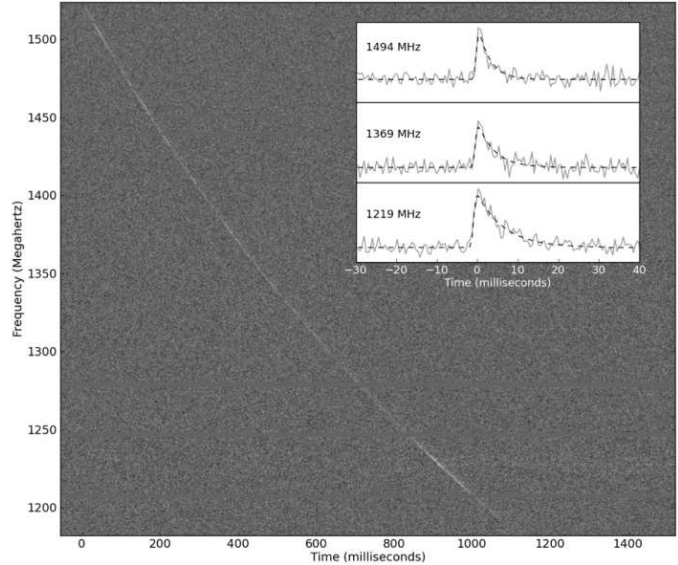,width=9cm}}
\vspace*{8pt}
\caption{One of the Thornton et al. (2013) FRBs.
    The telltale dispersion sweep of this single pulse is obvious across the radio band,
    and the burst profile at different radio frequencies is shown in the inset, where the
    radio-frequency dependence of the observed pulse broadening due to scattering is clear.
\label{fig:frb}}
\end{figure}

Finally, a newly discovered class of radio sources -- or rather, radio
events -- merits mention, even though they may or may not be related
to neutron stars.  Fast Radio Bursts (FRBs) are single, short (few
ms), bright (several Jy), highly dispersed radio pulses whose
dispersion measures suggest an origin far outside our Galaxy and
indeed at cosmological distances.\footnote{Note that FRBs are
  different in their properties from so-called ``perytons'', which
  turned out to be caused by local radio interference at the radio
  telescope site.\cite{pkb+15}}
  The first FRB reported\cite{lbm+07}
consisted of a single broadband radio burst lasting no longer than 5
ms from a direction well away from the Galactic Plane.  The burst was
extremely bright, with peak flux of 30 Jy, appearing for that moment
as one of the brightest radio sources in the sky.  The burst
dispersion measure was a factor of 15 times the expected contribution
from our Galaxy.  Thornton et al. (2013) reported\cite{tsb+13} 4 more
FRBs (see Fig.~\ref{fig:frb}), demonstrating the existence of a new
class of astrophysical events.  Concerns that FRBs could be an
instrumental phenomenon (since the Lorimer FRB and those reported by
Thornton were all found using the Parkes Observatory in Australia)
have recently been laid to rest by the discovery\cite{sch+14} of an
FRB using the Arecibo telescope. Another FRB discovered in
real-time\cite{pbb+14} was found to be $14-20$\%  circularly
polarised on the leading edge. No linear polarisation was detected,
although depolarisation due to Faraday rotation caused by passing through strong
magnetic fields and/or high-density environments cannot be ruled
out. The apparent avoidance of the Galactic Plane by FRBs is
consistent with a cosmological origin\cite{pvj+14} and an
event rate of $\sim 10^4$ per sky per day\cite{tsb+13}, a surprisingly large
number, albeit still based on small number statistics. Recent 
further data analysis and discoveries may suggest that this number may
be a little smaller but still consistent with the previously estimated
uncertainties (Champion, priv. comm.). 

One may wonder, why it took six years since the first ``Lorimer Burst'' to
discover further FRBs. This is due to the requirement to cover large
areas of the sky with sufficient time and frequency resolution,
combined with a need for sufficient computing power -- areas,
where recent modern surveys that are all based on digital hardware,
are superior to their predecessors.  Thus pulsar and RRAT hunters today are in
unique positions to find FRBs, in particular with new instruments
coming online that allow much larger fields-of-view.

The inferred large event rate and other FRB properties (DMs, widths,
the presence of a scattering tail in some cases; see
Fig.~\ref{fig:frb}) demand an explanation.  The locations on the sky
of the known FRBs are determined only to several arcminutes, a region
that typically contains many galaxies.  Hence identification of a host
galaxy -- key for understanding the nature of the burster and its
environment -- has been impossible.  Nevertheless, some models have
been proposed; papers in the refereed literature have appeared faster
than FRB detections! We discuss some of those models in Part II. with
reference of their importance to fundamental physics.

FRBs are thus highly reminiscent of the now-famous 
`Gamma Ray Burst' problem of the 1970s and 1980s -- sudden, unpredictable burst 
events on the sky and difficult to localize  -- though with FRBs having the added 
difficulty of dispersion and the attendant great delay in detection presently 
due to computational demands.  We cannot presently rule out that FRBs 
may represent a hitherto unrecognized type of astrophysical object, although as 
described below, neutron stars are also a plausible possibility.


\section*{Part II. - Neutron Stars as Laboratories for Fundamental Physics}

As described above, the vast majority of neutron stars have been discovered
in the radio regime in the form of radio pulsars. Putting aside astrophysical
population and pulsar emission issues, radio
observations of pulsars are important for totally independent reasons:
they add to other techniques and methods
employed to study fundamental physics with astronomical means. The latter
include the study of a possible variation of fundamental constants across cosmic time using 
molecular spectroscopy of emission that originates from distant
quasars. One can study the radio photons of the Cosmic Microwave
Background (CMB) in great detail, as is being done as part of this 
conference. One can also use the coherent emission of water maser
sources to obtain an accurate distance ladder to measure the local
expansion of the Universe. Table \ref{tab:radiotests} gives an
overview of such experiments with references for further reading.

\begin{table}[ph]
\tbl{Selected aspects of fundamental physics studied with
radio astronomical techniques compared to other methods. Note that some
solar system tests have better numerical precision but are derived in
weak gravitational field of the Solar System. In contrast, binary
pulsar limits may sometimes be less constraining in precision, but they
are derived for strongly self-gravitating bodies where deviations are
expected to be larger. References are given for more information or
further reading. For a
general review see Will (2014), and for pulsar-related limits see Wex
(2014).}
{\footnotesize
\begin{tabular}{p{3cm}p{4cm}p{4cm}l}\toprule
Tested phenomena & Method & Radio astronomy & Ref.\\
\noalign{\medskip}
\hline
\hline
\noalign{\medskip}
\multicolumn{2}{l}{Variation of fundamental constants:}  & & \\
\noalign{\smallskip}
 Fine structure constant ($e^2/\hbar c$) & Clock comparison, radio
active decays, limit depending on redshift, $< \sim 10^{-16}$ yr$^{-1}$ &
Quasar spectra, $< 10^{-16}$ yr$^{-1}$ & \refcite{wil14} \\
\noalign{\smallskip}
 e-p mass ratio & Clock comparison, $<3.3 \times 10^{-15}$ yr$^{-1}$ &
Quasar spectra, $<3 \times 10^{-15}$ yr$^{-1}$ & \refcite{blc+08},\refcite{ipv+05} \\
\noalign{\smallskip}
 Gravitational constant, $\dot{G}/G$ & Lunar Laser Ranging (LLR),  $(-0.7\pm
3.8) \times 10^{-13}$ yr$^{-1}$ & Binary pulsars, $(-0.6\pm 3.2) \times
10^{-12}$ yr $^{-1}$ & \refcite{hmb10,fwe+12,wex14}  \\
\noalign{\medskip}
\hline 
\noalign{\medskip}
 Universality of free fall: & LLR, Nordvedt parameter, $|\eta_N| = (4.4 \pm 
4.5) \times 10^{-4}$ &
Binary Pulsars, $\Delta < 5.6\times 10^{-3}$  &
\refcite{wil14,sfl+05,wex14} \\
\noalign{\medskip}
\hline 
\noalign{\medskip}
\multicolumn{2}{l}{Universal preferred frame for gravity:} & see Table
2 &  \\
\noalign{\medskip}
\hline 
\noalign{\medskip}
\multicolumn{2}{l}{PPN parameters and related phenomena:} & see Table
2 & \\
\noalign{\medskip}
\hline 
\noalign{\medskip} 
\multicolumn{2}{l}{Gravitational wave properties:}  &  Binary pulsars & \refcite{wex14} \\
\noalign{\smallskip}
Verification of GR’s quadrupole formula & & Double Pulsar,
$<3\times 10^{-4}$ & \refcite{ks+14} \\
\noalign{\smallskip}
Constraints on dipolar radiation &  & PSR-WD systems,  
$(\alpha_A - \alpha_B)^2 < 4 \times 10^{-6}$ & \refcite{fwe+12,afw+13}  \\
\noalign{\medskip}
\hline 
\noalign{\medskip} 
Geodetic precession & Gravity Probe B, 0.3\% & 
PSR B1913+16; Double Pulsar, 13\%; PSR B1534+12, 17\% &
\refcite{edp+11,kra98,bkk+08} \\
\noalign{\medskip}
\hline 
\noalign{\medskip}  
Equation-of-State & e.g. thermal emission from X-ray binaries & fast
spinning pulsars; massive neutron stars &
\refcite{ls14,hrs+06,dpr+10,afw+13,gswr13} \\
\noalign{\medskip}
\hline 
\noalign{\medskip}  
Cosmology & e.g. Supernova distances & CMB &
this conference \\
\noalign{\medskip}
\botrule
\end{tabular}}\label{tab:radiotests}
\end{table}

In the following, we will concentrate mostly on the study of 
gravitational physics where neutron star observations provide us with
the best tests and constraints existing todate.  Most of these tests
are possible due to the rotational stability of neutron stars;
the very large amount of stored 
rotational energy  ($\sim 10^{44}$ W), in particular that of 
the fast rotating millisecond pulsars, makes them effective
flywheels, delivering a radio ``tick'' per rotation with a precision
that rivals the best atomic clocks on Earth. At the same time they
are strongly self-gravitating bodies, enabling us to test not only the
validity of general relativity, but also to probe effects predicted by
alternative theories of gravity. They act as sources of gravitational
wave (GW) emission, if they are in a compact orbit with a binary
companion, but they may also act as detectors of low-frequency
GWs in a so-called ``pulsar timing array'' (PTA) experiment, as we discuss next.

\section{Tests of Theories of Gravity}

The idea behind the usage of pulsars for testing general relativity
(GR) and alternative theories of gravity is straightforward: if the
pulsar is in orbit with a binary companion, we use the measured
variation in the arrival times of the received signal to determine and
trace the orbit of the pulsar about the common centre of mass as the former
moves in the local curved space-time and in the presence of spin
effects. In alternative theories, self-gravity effects are often
expected, modifying also the orbital motion to be observed.

This ``pulsar timing'' experiment is simultaneously clean, conceptually
simple and very precise.  The latter is true since when measuring the
exact arrival time of pulses at our telescope on Earth, we do a
ranging experiment that is vastly superior in precision than a simple
measurement of Doppler-shifts in the pulse period. This is possible
since the pulsed
nature of our signal links tightly and directly to the rotation
of the neutron star, allowing us to count every single rotation. 
Furthermore, in this experiment we can consider the
pulsar as a test mass that has a precision clock attached to it.

While, strictly speaking, binary pulsars move in the weak
gravitational field of a companion, they do provide precision tests of
the (quasi-stationary) strong-field regime. This becomes clear when
considering that the majority of alternative theories predicts strong
self-field effects which would clearly affect the pulsars' orbital
motion. Hence, tracing their fall in a gravitational potential, we can
search for tiny deviations from GR, which can provide us with unique precision
strong-field tests of gravity.

\begin{table}[ph]
\tbl{Best limits for the parameters in the PPN formalism. 
Note that 6 of the 9 independent PPN parameters are best
constrained by radio astronomical techniques. Five
of them are derived from pulsar observations.
Adapted from Will (2014) but see also Wex (2014) for details.}
{\footnotesize
\begin{tabular}{lp{4cm}lrp{5cm}}\toprule
Par. & Meaning & Method  & Limit  & Remark/Ref.\\
\noalign{\medskip}
\hline
\hline
\noalign{\medskip}
$\gamma-1$ &  How much space-curvature produced by unit rest mass? & time delay &  $2.3 \times 10^{-5}$ & Cassini tracking/\refcite{wil14} \\
  & & light deflection &  $2\times 10^{-4}$ & VLBI/\refcite{wil14} \\
\noalign{\medskip}
\hline
\noalign{\medskip}
$\beta -1$ & How much ``non-linearity'' in the superposition law for
gravity? & perihelion shift & $8 \times 10^{-5}$ &  using
$J_{2\odot}=(2.2\pm0.1)\times 10^{-7}$/\refcite{wil14} \\
 &  & Nordtvedt effect &  $2.3\times 10^{-4}$ &  $\eta_{N}  =4\beta-\gamma-3$ assumed/\refcite{wil14} \\
\noalign{\medskip}
\hline
\noalign{\medskip}
$ \xi$ & Preferred-location effects? & spin precession & $4\times 10^{-9}$ & Isolated MSPs/\refcite{sw13} \\
\noalign{\medskip}
\hline
\noalign{\medskip}
$\alpha_1$ & Preferred-frame effects? & orbital polarisation  & $4\times 10^{-5}$ & PSR-WD, PSR J1738+0333/\refcite{sw12} \\
\noalign{\smallskip}
$\alpha_2$ &  & spin precession & $2 \times 10^{-9}$ & Using isolated MSPs/\refcite{sck+13} \\
\noalign{\smallskip}
$\alpha_3$ & & orbital polarisation  & $4\times 10^{-20}$ & Using ensemble
of MSPs/\refcite{sfl+05} \\
\noalign{\medskip}
\hline
\noalign{\medskip}
$\zeta_1$ & Violation of conservation of total momentum? & Combining PPN bounds &
$2\times 10^{-2}$ & \refcite{wil14} \\
\noalign{\smallskip}
$\zeta_2$ & & binary acceleration & $4\times 10^{-5}$ & Using $\ddot{P}$ for
PSR B1913+16/\refcite{wil14} \\
\noalign{\smallskip}
$\zeta_3$ & & Newton's 3rd law & $10^{-8}$ & lunar acceleration/\refcite{wil14} \\
\noalign{\smallskip}
$\zeta_4$ &  & not independent parameter  &  &  $6\zeta_4 = 3\alpha_3 + 2\zeta_1 − 3\zeta_3$  \\
\noalign{\medskip}
\botrule 
\end{tabular}}\label{tab:PPNpar}
\end{table}

As a result, a wide range of relativistic effects can be observed,
identified and studied.  These are summarised in
Table~\ref{tab:PPNpar} in the form of limits on the parameters in the
``Parameterised Post-Newtonian'' (PPN) formalism
(Ref.~\refcite{wil14}) and include concepts and principles deeply
embedded in theoretical frameworks.  If a specific alternative theory
is developed sufficiently well, one can also use radio pulsars to test
the consistency of this theory.  Table~\ref{tab:theories} lists a
number of theories where this has been possible. Sometimes, however,
gravitational theories are put forward to explain certain
observational phenomena without having studied the consequences of
these theories in other areas of parameter space. In particular,
alternative theories of gravity are sometimes proposed without having
worked out their radiative properties, while in fact,
tests for gravitational radiation provide a very powerful and
sensitive probe for the consistency of the theory with observational
data.  In other words, every successful theory has to pass the binary
pulsar experiments.

\begin{table}[ph]
\tbl{Constraining specific (classes of) gravity theories using radio pulsars.
See text and also Wex (2014) for more details.}
{\footnotesize
\begin{tabular}{lp{8cm}p{3cm}}\toprule
Theory (class)  & Method & Ref. \\
\noalign{\medskip}
\hline
\noalign{\medskip}  
\underline{Scalar-tensor gravity:} & & \\
\noalign{\smallskip}  
Jordan-Fierz-Brans-Dicke &  limits by PSR J1738+0333 and PSR
J0348+0432, comparable to best Solar system test (Cassini)  & \refcite{fwe+12} Freire priv. comm.  \\
\noalign{\smallskip}  
Quadratic scalar-tensor gravity &  for $\beta_0 < -3$ and $\beta_0 >
0$ best limits from PSR-WD systems, in particular PSR J1738+0333 and
PSR J0348+0432 &  \refcite{fwe+12} Krieger et al.~in prep., Freire priv. comm.  \\
\noalign{\smallskip}   
Massive Brans-Dicke & for $m_\varphi \sim 10^{-16}$ eV: PSR J1141$-$6545 &  \refcite{abwz12}\\
\noalign{\medskip}  
\hline  
\noalign{\medskip}   
\underline{Vector-tensor gravity:} & & \\
\noalign{\smallskip}    
Einstein-\AE{}ther & combination of pulsars (PSR J1141$-$6545, 
PSR J0348+0432, PSR J0737$-$3039, PSR J1738+0333) &  \refcite{ybby14}\\
\noalign{\smallskip}  
Ho{\v r}ava gravity & combination of pulsars (see above) & \refcite{ybby14}   \\
\noalign{\medskip}  
\hline 
\noalign{\medskip}  
\underline{TeVeS and TeVeS-like theories:} & & \\
\noalign{\smallskip}    
Bekenstein’s TeVeS & excluded using Double Pulsar  &  \refcite{ks+14} \\
\noalign{\smallskip}   
TeVeS-like theories & excluded using PSR 1738+0333 & \refcite{fwe+12} \\
\noalign{\medskip}
\botrule 
\end{tabular}}\label{tab:theories}
\end{table}

The various effects or concepts  to be tested require sometimes rather
different  types of  laboratory. For  instance, in  order to  test the
important radiative properties  of a theory, we  need compact systems,
usually consisting of a pair of neutron stars. As we have seen, double
neutron  star systems  (DNSs) are  rare but  they usually  produce the
largest observable  relativistic effects in their  orbital motion and,
as  we will  see, produce  the best  tests of  GR  for
strongly  self-gravitating bodies.   On the  other hand,  to test  the
violation of the Strong Equivalence Principle, one would like to use a
binary system  that consists of  different types of masses  (i.e. with
different  gravitational self-energy),  rather than  a system  made of
very similar bodies, so that we  can observe how the different masses
fall in  the gravitational  potential of the  companion and of the Milky
Way.  For  this  application,  a pulsar-black  hole  system  would  be
ideal. Unfortunately, despite  past and ongoing efforts,  we have not
yet found a  pulsar orbiting a stellar black hole  companion or orbitting the
supermassive   black    hole   in    the   centre   of    our   Galaxy
\cite{lwk+12}. Fortunately, we can use pulsar-white dwarf (PSR-WD)
systems, as white dwarfs and neutron stars differ very significantly in their 
structure and, consequently, self-energies. Furthermore, some PSR-WD
systems can also be found in relativistic orbits\cite{klm+00a,lbr+13}.

\section{The First Binary Pulsar -- a Novel Gravity Laboratory}

The first binary pulsar to ever be discovered happend to be a rare
double neutron star system. It was discovered by Russel Hulse and Joe
Taylor in 1974 (Ref.~\refcite{ht75a}). The pulsar, B1913+16, has a period of
59\,ms and is in an eccentric ($e=0.62$) orbit around an unseen companion
with an orbital period of less than 8 hours. Soon after the discovery,
Taylor and Hulse noticed that
the pulsar does not follow the movement expected from a simple
Keplerian description of the binary orbit, but that it shows the
impact of relativistic effects.  In order to describe the relativistic
effects in a theory-independent fashion, one introduces so-called
``Post-Keplerian'' (PK) parameters that are included in a timing model
to describe accurately the measured pulse times-of-arrival (see
e.g.~Ref.~\refcite{lk05} for more details).

For the Hulse-Taylor pulsar, 
a relativistic advance of its periastron was soon measured analogous to what is seen in the solar system for Mercury, albeit
with a much larger amplitude. The value measured today, $\dot\omega = 4.226598 \pm 0.000005$
deg/yr \cite{wnt10}, is much more precise than than was originally measured, but even early on the
precision
was sufficient to permit meaningful comparisons with GR's prediction. The value depends on
the Keplerian parameters and the masses of the pulsar and its
companion:
\begin{equation}
\dot{\omega} = 3 T_\odot^{2/3} \; \left( \frac{\pb}{2\pi} \right)^{-5/3} \;
               \frac{1}{1-e^2} \; (\mpp +
               \mcc)^{2/3}. \label{omegadot}
\end{equation}
Here, $T_\odot=GM_\odot/c^3=4.925490947 \mu$s is a
constant, $\pb$ the orbital period, $e$ the
eccentricity, and $\mpp$ and $\mcc$ the masses of the pulsar and its
companion, respectively. See Ref.~\refcite{lk05} for further details.

The Hulse-Taylor pulsar also shows the effects of gravitational redshift (including
a contribution from a second-order Doppler effect) as the pulsar moves
in its elliptical orbit at varying distances from the companion and
with varying speeds.  The result is a variation in the clock rate of
with an amplitude of $\gamma = 4.2992 \pm 0.0008$ ms (Ref.~\refcite{wnt10}). In GR, the
observed value is related to the Keplerian parameters and the masses as
\begin{equation}
\gamma  = T_\odot^{2/3}  \; \left( \frac{\pb}{2\pi} \right)^{1/3} \;
              e\frac{\mcc(\mpp+2\mcc)}{(\mpp+\mcc)^{4/3}}.
\end{equation}
We can now combine these measurements. We have two equations with a
measured left-hand side. On the right-hand side, we measured
everything apart from two unknown masses. We solve for those and
obtain, $\mpp = 1.4398 \pm 0.0002 \,M_\odot$ and $\mcc = 1.3886 \pm 0.0002
\,M_\odot$ \cite{wnt10}.  These masses are correct if GR is the right theory of
gravity. If that is indeed the case, we can make use of the fact that
(for point masses with negligible spin contributions), the PK
parameters in each theory should only be functions of the {\it a priori}
unknown masses of pulsar and companion, $\mpp$ and $\mcc$, and the
easily measurable Keplerian parameters (Ref.~\refcite{dt91})\footnote{For
  alternative theories of gravity this statement may only be true for
  a given equation-of-state.}.  With the two masses now being
determined using GR, we can compare any observed value of a third PK
parameter with the predicted value. A third such parameter is the
observed decay of the orbit which can be explained fully by the emission
of gravitational waves. And indeed, using the derived masses, along with the
prediction of GR, i.e.
\begin{equation}
\dot{P}_{\rm b} = -\frac{192\pi}{5} T_\odot^{5/3} \; \left( \frac{\pb}{2\pi} \right)^{-5/3} \;
               \frac{\left(1 +\frac{73}{24}e^2 + \frac{37}{96}e^4 \right)}{(1-e^2)^{7/2}} \; 
               \frac{\mpp \mcc}{(\mpp + \mcc)^{1/3}},
\end{equation}
one finds an agreement with the observed value of 
$\dot{P}_{\rm b}^{\rm obs} = (‐2.423 \pm 0.001) \times 10^{-12}$ 
(Ref.~\refcite{wnt10}) --
however, only if a correction for a relative acceleration between the
pulsar and the solar system barycentre is taken into account. As
the pulsar is located about 7 kpc away from Earth, it experiences a
different acceleration in the Galactic gravitational potential than
does the solar system (see e.g. ~Ref.~\refcite{lk05}). The precision of our knowledge to correct for this
effect eventually limits our ability to compare the GR prediction to
the observed value. Nevertheless, the agreement of observations and
prediction, today within a 0.2\% (systematic) uncertainty\cite{wnt10}, 
represented the first evidence for the existence of
gravitational waves. Today we know many more binary pulsars in which  
we can detect the effects of gravitational wave emission. In one particular case, the
measurement uncertainties are not only more precise, but also the
systematic uncertainties are much smaller, as the system is much more
nearby. This system is the Double Pulsar.

\section{The Double Pulsar}

The Double Pulsar was discovered in 2003\cite{bdp+03,lbk+04}. It
not only shows larger relativistic effects and is much closer to
Earth (about 1 kpc) than the Hulse-Taylor pulsar, allowing us to
largely neglect the relative acceleration effects, but the defining unique
property of the system is that it does not consist of one active
pulsar and its {\em unseen} companion, but that it harbours two {\em
  active} radio pulsars.

One pulsar is mildly recycled with a period of 23 ms (named ``A''), while the other
pulsar is young with a period of 2.8 s (named ``B''). Both orbit the common centre
of mass in only 147-min with orbital velocities of 1 Million km per
hour. Being also mildly eccentric ($e=0.09$), the system is an ideal
laboratory to study gravitational physics and fundamental physics in
general. A detailed account of the exploitation for gravitational
physics has been given, for instance, by Refs.~\refcite{ksm+06,ks08,kw09}. 
An update on those results is in preparation\cite{ks+14},
 with the largest improvement undoubtedly given by a
large increase in precision when measuring the orbital decay. Not
even ten years after the discovery of the system, the Double Pulsar
provides the best test for the accuracy of the gravitational
quadrupole emission prediction by GR far below the 0.1\% level.

\begin{figure}
\centerline{\psfig{file=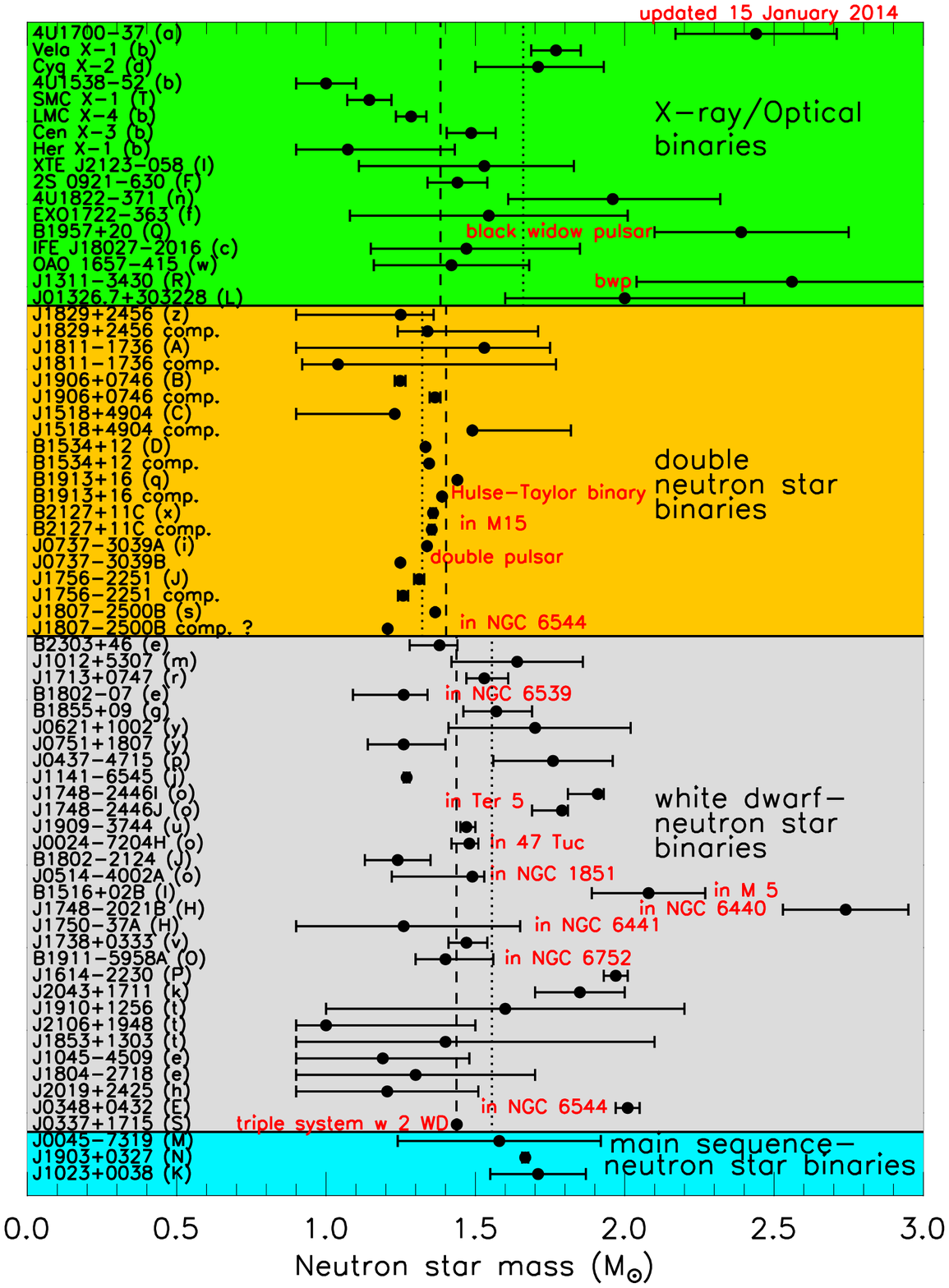,width=11cm}}
\caption{Neutron star mass measurements compiled by J.~Lattimer and
  available at {\tt  www.stellarcollapse.org}.
\label{fig:masses}}
\end{figure}

In order to perform this test, we first determine the mass ratio of
pulsar A and B from their relative sizes of the orbit, i.e.~$R=x_B/x_A
= m_A / m_B = 1.0714\pm 0.0011$ \cite{ksm+06}. Note that this
value is theory-independent to the 1PN level\cite{dd86}.
The most precise PK parameter that can be measured is a large orbital
precession, i.e.~$\dot\omega = 16.8991 \pm 0.0001$ deg/yr. 
Using Eq.~(\ref{omegadot}), this measured value and the mass ratio,
we can determine the masses of the pulsars, assuming GR is correct,
to be $m_A = (1.3381\pm0.0007) \,M_\odot$ and $m_B= (1.2489\pm0.0007)
\,M_\odot$. The masses are shown, together with others determined by
this and other methods, in Figure~\ref{fig:masses}.

We can use these masses to compute the expected amplitude for the
gravitational redshift, $\gamma$, if GR is correct. Comparing the result
with the observed value of $\gamma = 383.9 \pm 0.6~\mu$s, we find that 
theory (GR) agrees with the observed value to a ratio of
$1.000\pm0.002$, as a first of five tests of GR in the Double Pulsar.

The Double Pulsar also has the interesting feature that the orbit is 
seen nearly exactly edge-on. This leads to a 30-s long eclipse of
pulsar A due to the blocking magnetosphere of B that we discuss further below,
but it also leads to a ``Shapiro delay'': whenever the pulse needs to
propagate through curved space-time, it takes a little longer than
travelling through flat space-time. At superior conjunction, when
the signal of pulsar A passes the surface of B in only 20,000~km
distance, the extra path length due to the curvature of space-time
around B leads to an extra time delay of about 100~$\mu$s. The
shape and amplitude of the corresponding Shapiro delay curve yield two
PK parameters, $s$ and $r$, known as {\em shape} and {\em range},
allowing two further tests of GR.
$s$ is measured to $s = \sin(i)=0.99975 \pm 0.00009$ and is in
agreement with the GR prediction of
\begin{equation}
s = T_\odot^{-1/3} \; \left( \frac{\pb}{2\pi} \right)^{-2/3} \; x \;
              \frac{(m_A + m_B)^{2/3}}{m_B} \label{eqn:s},
\end{equation}
(where $x$ is the projected size of the semi-major axis measured in lt-s)
 within a ratio of $1.0000\pm 0.0005$.
It corresponds to an orbital inclination angle of $88.7\pm0.2$ deg,
which is indeed very close to 90 deg as suggested by the eclipses.
$r$ can be measured with much less precision and yields an agreement
with GR's
value given by
\begin{equation}
r = T_\odot m_B, \label{eqn:r}\\
\end{equation}
to within a factor of $0.98\pm0.02$.

A fourth test is given by comparing an observed orbital decay of
$107.79\pm0.11$ ns/day to the GR prediction. Unlike the Hulse-Taylor
pulsar, extrinsic effects are negligible and the values agree with
each other without correction to within a ratio of
$1.000\pm0.001$. This is already a better test for the existence of GW
than possible with the Hulse-Taylor pulsar and will continue to improve
with time. Indeed, at the time of writing the
agreement has already surpassed the 0.03\% level\cite{ks+14}.

\section{Relativistic Spin-orbit Coupling}

Apart from the Shapiro-delay, the impact of curved space-time is also
immediately measurable by its effect on the orientation of the pulsar
spin in a gyroscope experiment. This effect, known as geodetic
precession or de Sitter precession represents the effect on a vector
carried along with an orbiting body such that the vector points in a
different direction from its starting point (relative to a distant
observer) after a full orbit around the central
object. Experimental verification has been achieved
by precision tests in the solar system, e.g. by Lunar Laser Ranging
(LLR) measurements, or recently by measurements with the Gravity
Probe-B satellite mission (see Table~\ref{tab:radiotests}).
However, these tests are done in the weak field
conditions of the solar system.  Thus Pulsars currently provide the only
access beyond weak-field, i.e.~the quasi-stationary strong-field regime.

In binary systems one can interpret the observations, depending on the
reference frame, as a mixture of different contributions to
relativistic spin-orbit interaction. One contribution comes from the
motion of the first body around the centre of mass of the system
(de Sitter-Fokker precession), while the other comes from the dragging
of the internal frame at the first body due to the translational
motion of the companion\cite{ber75}.
Hence, even though we loosely talk about
geodetic precession, the result of the spin-orbit coupling for binary
pulsars is more general, and hence we will call it {\em relativistic
  spin-precession}. The consequence of relativistic spin-precession is a precession of
the pulsar spin about the total angular moment vector, changing the
orientation of  the pulsar relative to Earth.

Since the orbital angular momentum is much larger than the spin of
the pulsar, the orbital angular momentum practically represents a
fixed direction in space, defined by the orbital plane of the binary
system. Therefore, if the spin vector of the pulsar is misaligned with the
orbital spin, relativistic spin-precession leads to a change in
viewing geometry, as the pulsar spin precesses about the total angular
momentum vector.  Consequently, as many of the observed pulsar
properties are determined by the relative orientation of the pulsar
axes towards the distant observer on Earth, we should expect a
modulation in the measured pulse profile properties, namely its shape
and polarisation characteristics\cite{dr74}.
The precession rate is another PK parameter and given in GR by
(e.g.~Ref.~\refcite{lk05})
\begin{equation}
\label{eqn:om}
\Omega_{\rm p}  =  T_\odot^{2/3} \left( \frac{2\pi}{P_{\rm b}}\right)^{5/3} 
 \frac{m_{\rm c}(4m_{\rm p}+3m_{\rm c})}{2(m_{\rm p}+m_{\rm c})^{4/3}} 
 \frac{1}{1-e^2}.
\end{equation}
In order to see a measurable effect in any binary pulsar, {\em a)} the spin
axis of the pulsar needs to be misaligned with the total angular
momentum vector and {\em b)} the precession rate must be sufficiently
large compared to the available observing time to detect a change in
the emission properties. Considering these conditions, relativistic
spin precession has now been detected in {\em all} systems where we
can realistically expect this.

As the most relativistic binary system known to date, we expect a
large amount of spin precession in the Double Pulsar system.  Despite
careful studies, profile changes for A have not been detected,
suggesting that A's misalignment angle is less than a few degrees
\cite{fsk+13}.  In contrast, changes in the light curve and pulse
shape on secular timescales\cite{bpm+05} reveal that this is not the
case for B. In fact, B had been becoming progressively weaker and
disappeared from our view in 2009\cite{pmk+10}. Making the 
assumption that this disappearance is solely caused by relativistic
spin precession, it will only be out of sight temporarily until it
reappears later. Modelling suggests that, depending on the beam shape,
this will occur in about 2035  but an earlier time cannot be
excluded.\cite{pmk+10}. The geometry that is derived from this modelling is
consistent with the results from complementary observations of spin
precession, visible via a rather unexpected effect described in the
following.

The change on the orientation of B also changes the observed eclipse
pattern in the Double Pulsar, where we can see periodic bursts of
emission of A during the dark eclipse phases, with the period being
the full- or half-period of B. As this pattern is caused by the
rotation of B's blocking magnetospheric torus that allows light to
pass B when the torus rotates to be seen from the side, the resulting
pattern is determined by the three-dimensional orientation of the
torus, which is centred on the precessing pulsar spin. Eclipse
monitoring over the course of several years shows exactly the expected
changes, allowing a determination of the precession rate to $\Omega_{\rm p,
  B}= 4.77^{+0.66}_{-0.65}$ deg/yr.  This value is fully consistent
with the value expected in GR, providing a fifth test\cite{bkk+08}.
This measurement also allows us to test alternative theories of gravity and
their prediction for relativistic spin-precession in strongly
self-gravitating bodies for the first time (see Ref.~\refcite{kw09} for
details).

\section{Alternative Theories of Gravity}

Despite the successes of GR, a range of observational data has
fuelled the continuous development of alternative theories of gravity.
Such data include the apparent observation of ``dark matter'' or the
cosmological results interpreted in the form of ``inflation'' and
``dark energy,'' as also discussed at this conference. Confronting
alternative theories with data also in other areas of the parameter
space (away from the CMB or Galactic scales), requires that these
theories are developed sufficiently in order to make predictions. As
mentioned, a particularly sensitive criterion is if the theory is able
to make a statement about the existence and type of gravitational
waves emitted by binary pulsars. Most theories cannot do this (yet),
but a class of theories where this has been achieved is the class of
tensor-scalar theories as discussed and demonstrated by Damour and
Esposito-Far{\`e}se in a series of works (e.g.~Ref.~\refcite{de96}).
For corresponding tests, the choice of a double neutron star system is
not ideal, as the difference in scalar coupling, (that would be
relevant, for instance, for the emission of gravitational {\em dipole}
radiation) is small. The ideal laboratory would be a pulsar orbiting a
black hole, as the black hole would have zero scalar charge.
The next best laboratory is a
pulsar-white dwarf system. Indeed, such binary systems are able to
provide constraints for alternative theories of gravity that are
equally good or even better than solar system limits\cite{fwe+12}.

The previously best example for such a system was presented by
Ref.~\refcite{fwe+12}, who reported the results of a 10-year timing
campaign on PSR J1738+0333, a 5.85-ms pulsar in a practically circular
8.5-h orbit with a low-mass white dwarf companion. A large number of
precision pulse time-of-arrival measurements allowed the determination
of the intrinsic orbital decay due to gravitational wave emission.
The agreement of the observed value with the prediction of GR
introduces a tight upper limit on dipolar gravitational wave emission,
which can be used to derive the most stringent constraints ever on
general scalar-tensor theories of gravity. The new bounds are more
stringent than the best current Solar system limits over most of the
parameter space, and constrain the matter-scalar coupling constant
$\alpha_0^2$ to be below the $10^{-5}$ level. For the special case of
the Jordan-Fierz-Brans-Dicke theory, the authors obtain a one-sigma
bound of $\alpha_0^2< 2 \times 10^{-5}$, which is within a factor of
two of the Solar-System Cassini limit.\cite{fwe+12,bit03}
Moreover, their limit on
dipolar gravitational wave emission can also be used to constrain a
wide class of theories of gravity which are based on a generalisation
of Bekenstein's Tensor-Vector-Scalar (TeVeS) gravity, a relativistic
formulation of Modified Newtonian Dynamics (MOND)\cite{bek04}. They find that in
order to be consistent with the results for PSR J1738+0333, these
TeVeS-like theories have to be fine-tuned significantly 
(see Table~\ref{tab:theories}). We expect the latest
Double Pulsar results to close a final gap of parameter spaced left
open by the PSR-WD systems.\cite{fwe+12,ks+14}

A recently studied pulsar-white dwarf system \cite{lbr+13,afw+13}
turned out to be a very exciting laboratory for various aspects of
fundamental physics: PSR J0348+0432 harbours a white dwarf whose
composition and orbital motion can be precisely derived from optical
observations.  The results allow us to measure the mass of the neutron
star, showing that it has a record-breaking value of
$2.01\pm0.04M_\odot$!\cite{afw+13} This is not only the most massive
neutron star known (at least with reliable precision), providing
important constraints on the ``equation-of-state'' (see below) but the
39-ms pulsar and the white dwarf orbit each other in only 2.46 hours,
i.e. the orbit is only 15 seconds longer than that of the Double
Pulsar. Even though the orbital motion is nearly circular, the effect
of gravitational wave damping is clearly measured. Hence, the high
pulsar mass and the compact orbit make this system a sensitive
laboratory of a previously untested strong-field gravity regime. Thus
far, the observed orbital decay agrees with GR, supporting its
validity even for the extreme conditions present in the
system\cite{afw+13}.  The precision of the observed agreement is
already sufficient to add significant confidence to the usage of GR
templates in the data analysis for gravitational wave (GW) detectors.

\section{Pulsars as Gravitational Wave Detectors}

The observed orbital decay in binary pulsars detected via precision
timing experiments so far offers the best evidence for the existence
of gravitational wave (GW) emission. Intensive efforts are therefore
ongoing world-wide to make a direct detection of gravitational waves
that pass over the Earth. Ground-based detectors like GEO600, VIRGO or
LIGO use massive mirrors, the relative separations of which are measured by a
laser interferometer set-up, while the envisioned space-based LISA
detector uses formation flying of three test-masses that are housed in
satellites. For a summary of these efforts, see, e.g.~Ref.~\refcite{wil14}.

The change of the space-time metric around the Earth also influences
the arrival times of pulsar signals measured at the
telescope. Therefore, pulsars do not only act as sources of GWs, but
they may eventually also lead to their direct detection.
Fundamentally, the GW frequency range that pulsar timing is sensitive
to, is bound by the cadence of the timing observations on the high
frequency side, and by the length of the data set on low-frequency
part. Hence, typically GWs with periods of the order of one year or
more could be detected. Since GWs are expected to produce a
characteristic quadrupole signature on the sky, the timing residuals
from various pulsars should be correlated correspondingly\cite{hd83},
so that the comparative timing of several pulsars can be used to make
a detection. The sensitivity of such a ``Pulsar Timing Array'' (PTA)
increases with the number of pulsars and should be able to detect
gravitational waves in the nHz regime, hence below the frequencies to
which LIGO ($\sim$kHz and higher) and LISA ($\sim$mHz) are
sensitive. Sources in the nHz range (see, e.g.,~Ref.~\refcite{jhm+15})
include astrophysical objects (i.e.~super-massive black hole binaries
resulting from galaxy mergers in the early Universe), cosmological
sources (e.g.~the vibration of cosmic strings), and transient
phenomena (e.g. phase transitions).

\begin{figure}
\centerline{\psfig{file=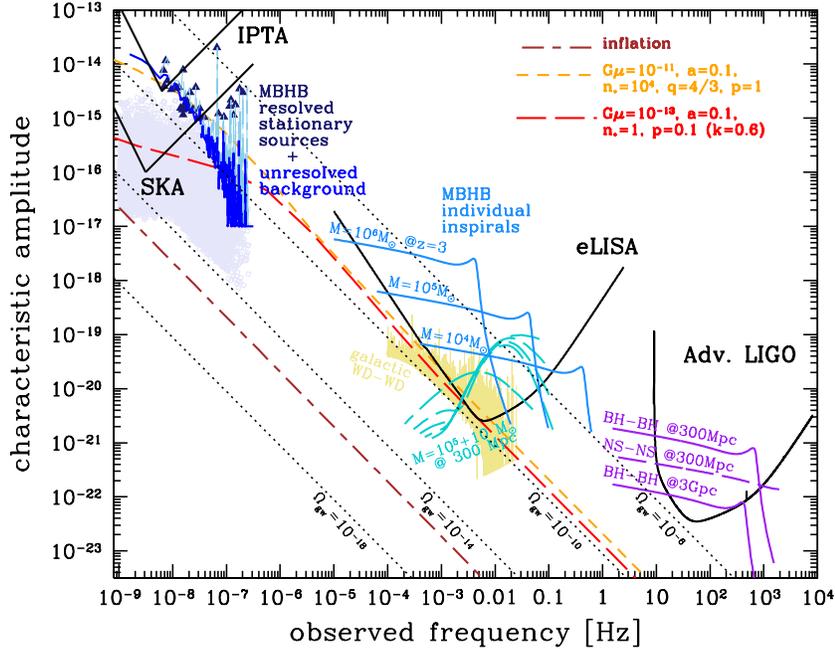,width=11cm}}
\caption{The gravitational wave spectrum with expected sources. Shown
  is the characteristic amplitude vs. frequency as presented by
  Janssen et al. (2015): In the nHz regime, individually resolvable
  systems and the level of the unresolved background are indicated.
Nominal sensitivity levels for the IPTA and SKA are also shown. In the mHz frequency range, the eLISA
sensitivity curve is shown together with typical circular SMBHB inspirals at z=3 (pale blue), the overall
signal from Galactic WD-WD binaries (yellow) and an example of extreme
mass ratio inspiral (aquamarine).
In the kHz range an advanced LIGO curve is shown together with selected compact object inspirals (purple). The brown, red
and orange lines running through the whole frequency range are expected cosmological backgrounds from
standard inflation and selected string models, as labeled in figure.
\label{fig:spectrum}}
\end{figure}

A number of PTA experiments are ongoing, namely in Australia, Europe
and North America (see~Ref.~\refcite{man13} for a summary). The currently
derived upper limits on a stochastic GW background
(e.g.~Refs.~\refcite{src+13}) are very close to the theoretical
expectation for a signal that originates from binary supermassive
black holes expected from the hierarchical galaxy evolution model
\cite{svc08,sv10}. 

But the science that can eventually be done with the PTAs goes far
beyond simple GW detection -- a whole realm of astronomy and
fundamental physics studies will become possible.  The dominant signal
in the nHz regime is expected to be a stochastic background due to
merging supermassive black holes and many constraints can be
placed on this source population, including their frequency in cosmic
history, the relation between the black holes and their hosts, and
their coupling with the stellar and gaseous environments
\cite{ses13,ses15}.  Detection of gas disks surrounding merging
supermassive black holes and related eccentricities in such systems is
possible \cite{rds+11,rs12}; PTAs should be able to constrain the
solution to the famous `last parsec problem' \cite{scm15}.  In
addition to detecting a {\em background} of GW emission, PTAs can
detect {\em single} GW sources.  We can, for instance, expect to
detect anisotropies in a GW background, due to the signals of single
nearby supermassive black hole binaries
\cite{lwk+11,msmv13}. Considering the case when the orbit is
effectively not evolving over the observing span, we can show that, by
using information provided by the ``pulsar term'' (i.e.~the retarded
effect of the GW acting on the pulsar's surrounding spacetime), we may
be able to achieve interesting ($\sim$1~arcmin) source localisation
\cite{lwk+11}.  Even astrophysical measurements of more local
relevance can be done with PTAs; for example an independent
determination of the masses of the Jovian planetary system has already
been made (Ref.~\refcite{chm+10}) and additional future, improved
measurements for Jupiter and other planets should be possible.  On the
fundamental physics side, departures from GR during
supermassive black hole mergers should be measurable via different
angular dependences of pulsar timing residuals on the sky such as for
example from gravitational wave polarization properties that differ
from those predicted by GR \cite{ljp08,cs12,mgs+12}.
It may even be possible to constrain the mass of the graviton from the
angular correlation of pulsar timing residuals \cite{ljp+10}.  If the
ongoing PTA experiments do not detect GWs in the next few years, a
first detection is virtually guaranteed with the more sensitive Phase
I of the Square Kilometer Array \cite{jhm+15}.  With even further
increased sensitivity of SKA Phase II, it should also be possible to
study the fundamental properties of gravitational waves.

\section{Black Holes or the Centre of the Galaxy as a Gravity Lab}

What makes a binary pulsar with a black-hole companion so interesting
is that it has the potential of providing a superb new probe of
relativistic gravity. As pointed out by Ref.~\refcite{de98}, the discriminating power of this probe might
supersede all its present and foreseeable competitors. The reason lies
in the fact that such a system would clearly expose the self-field
effects of the body orbiting the black hole, hence making it an
excellent probe for alternative theories of gravity.

But also for testing the black hole properties predicted by GR, a
pulsar-BH system will be superb laboratory.  
Ref.~\refcite{wk99} was the first to provide a detailed recipe for how
to exploit a pulsar-black hole system. They showed that the
measurement of spin-orbit coupling in a pulsar-BH binary in
principle allows us to determine the spin and the quadrupole moment
of the black hole.  This could test the ``cosmic censorship
conjecture'' and the ``no-hair theorem''. While Ref.~\refcite{wk99}
showed that with current telescopes such an experiment would be almost
impossible to perform (with the possible exception of pulsars about
the Galactic centre black hole), Ref.~\refcite{kbc+04}
pointed out that the SKA sensitivity should be sufficient. Indeed,
this experiment benefits from the SKA sensitivity in multiple ways. It
provides the required timing precision while also enabling
deep searches, enabling a Galactic Census which should eventually deliver the
desired sample of pulsars with a BH companion. As shown recently\cite{lewk14},
with the SKA or the {\em Five-hundred-meter Aperture Spherical radio
  Telescope} (FAST) project\cite{nlj+11} one could test the
cosmic censorship conjecture by measuring the spin of a stellar black
hole, though it is still unlikely to find a system that can enable
the measurement of the quadrupole moment.

As the effects become easier to measure with more massive black holes,
the best laboratory would be a pulsar orbiting the central black hole
in the Milky Way, Sgr A* \cite{wk99,pl04,kbc+04}. Indeed,
Ref.~\refcite{lwk+12} continued the work of Ref.~\refcite{wk99} and
studied this possibility in detail. They showed that it should be
``fairly easy'' to measure the spin of the GC black hole with a
precision of $10^{-4} - 10^{-3}$. Even for a pulsar with a timing
precision of only 100~$\; \mu$s, characteristic periodic residuals
would enable tests of the no-hair theorem with with a precision of one
percent or better!

\subsection{Pulsars in the Galactic Centre}

Unfortunately, searches for pulsars near Sgr A* have been unsuccessful
for the last 30 years - until April 2013. As described in
Section~\ref{sec:1745}, a radio signal of the 3.7-s magnetar
J1745$-$2900 was detected\cite{efk+13,sj13}.  The source has the
highest dispersion measure of any known pulsar, is highly polarised and has
a rotation measure that is larger than that of any other source in the
Galaxy, apart from Sgr A*. This, and the fact that VLBI images of the
magnetar show scattering identical to that in the radio image of Sgr A* itself
\cite{bdd+14}, support the idea that the source is indeed only
$\sim 0.1$ pc away from the central black hole. Initially,
measurements of the scatter-broadening of the single radio pulses\cite{sle+14} 
suggested that scattering due to the inner
interstellar medium is too small to explain the lack of pulsar
detection in previous survey. Recent preliminary results, enabled by
the the puzzling fact that the radio emission remains unabated in spite
of significant source fading in the X-ray band\cite{laks15}, show an
increase of scattering, indicating that the conditions are instead
highly changeable (Spitler et al.~in prep.). The fact that a rare
object like a radio-emitting magnetar is found in such proximity to
Sgr A* suggests that estimates like that of Ref.~\refcite{wcc+12},
predicting as many as 1000 pulsars in the inner central parsec, may
indeed be true. Further searches are ongoing but may require
observations at very high (i.e.~ALMA) frequencies, i.e.~$>40$ GHz to
beat the extreme scattering, which decreases as $\sim \nu^{-4}$.

\subsection{The Event Horizon Telescope \& BlackHoleCam}

Telescopes operating at high radio frequencies may not only allow us
to find a pulsar in the Galactic Centre, but combined with other radio
telescopes, they can also form an interferometer to take an image of
Sgr A* that can resolve the ``shadow'' of the supermassive black hole
in the centre of our Milky Way. With a mass of about $4.3\times 10^6
M_\odot$ \cite{gsw+08,gef+09} it is not very large in size compared to
those in the centres of other galaxies, but it is the closest. The
image to be taken by the so-called ``Event Horizon Telescope'' and
``BlackHoleCam'' experiments (see e.g. Ref.~\refcite{fm13} for a
recent review) will depend on the magnitude and direction of Sgr A*'s
spin, i.e.~information available by the discovery of pulsar around the
central black hole, as described above. Combined measurements probe
simultaneously the near- and far-field of Sgr A*, promising a unique
probe of gravity.

\section{Physics at Extreme Densities}

The density of pulsars and neutron stars is so large that their matter
cannot be reproduced in terrestrial observatories. Therefore, in order
to understand how matter behaves under very extreme condition,
observations of pulsars provide unique insight. On one hand, mass
measurements constrain the Equation-of-State (EOS) at the highest
densities, which also affects the maximum possible spin frequency of
pulsars\cite{hrs+06} and sets bounds to the highest possible density
of cold matter (see contribution by J.~Lattimer). Because, a given EOS
describes a specific mass-radius relationship
(see, e.g., Ref.~\refcite{poc14}), measurements of the radii of neutron
stars also set constrains on the EOS near nuclear saturation density
and yield information about the density dependence of the nuclear
symmetry energy\cite{gswr13,lat14}. In practice, mass measurements are easier
to achieve than radius measurements -- or the discovery of
sub-millisecond pulsars with significantly faster spin-periods than
currently known\cite{hrs+06}. Specifically, while there are about 40
neutron star masses known with varying accuracy (see
Figure~\ref{fig:masses}), there are no precise simultaneous
measurements of mass and radius for any neutron star.

\begin{figure}
\centerline{\psfig{file=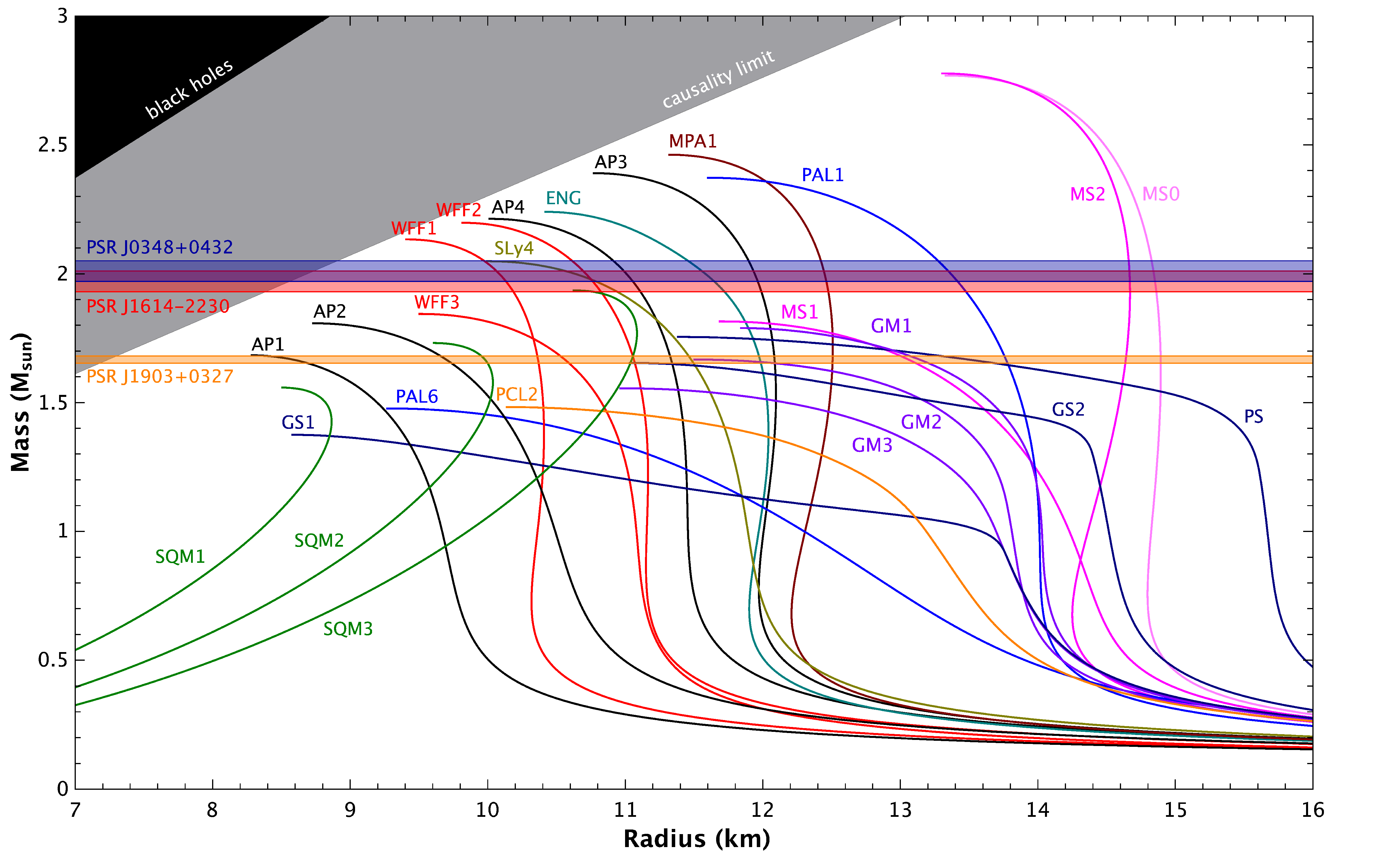,width=13cm}}
\caption{Constraints on the equation-of-state provided by mass
  measurements of the most massive neutron stars. Figure provided by
  N. Wex. For details see e.g.~Demorest et al.~(2010).
\label{fig:eos}}
\end{figure}

For now, some of the best constraints for the EOS come simply from the
maximum observed neutron star mass. Unlike in Newtonian physics, in GR
a maximum mass exists as for any causal EOS as the isothermal speed of
sound must never exceed the speed of light. Currently, the largest
masses are measured for PSR J1614+2230 with $M=1.94\pm0.04 M_\odot$
\cite{dpr+10} and PSR J0348+0432 with $M=2.01\pm0.04 M_\odot$
\cite{afw+13}.  These independent measurements confirm the existence
of high-mass neutron stars, ruling out a number of soft EOS already
(see Figure~\ref{fig:eos}). However, as explained, for instance, in
Ref.~\refcite{lat14}, this lower limit on the maximum mass also
provides constraints on the EOS at lower densities and on the radii of
intermediate mass neutron stars. In general, however, most radii
estimates come from estimates inferred from photospheric radius
expansion bursts and thermal X-ray emission from neutron star
surfaces. A Bayesian analysis of the existing data suggests a radius
range of 11.3--12.1 km for a $1.4M_\odot$ neutron star.\cite{gswr13,lat14}

In terms of information about fundamental properties of super-dense
matter, the {\em maximum} mass of neutron stars is clearly important. 
Small mass measurements, in particular those below $1.20M_\odot$, are
nevertheless extremely interesting from a neutron-star formation
point-of-view as they would call into question the gravitational-collapse
formation scenario\cite{lat14}. One way to form such
low-mass neutron stars is through electron-capture supernovae. Here, 
a white dwarf with an oxygen-neon-magnesium (O-Ne-Mg) core collapses 
to a low-mass neutron star due to electron captures on Ne and/or Mg,
as was proposed for the formation of the light companion in the
Double Pulsar system, PSR J0737$-$3039B\cite{pdl+05}. It was 
suggested that electron capture could be triggered in
particular in close binaries. Assuming minimal mass loss, the final 
mass should be determined by the mass of the progenitor star minus
the binding energy. As for any given EOS one can calculate the
relation between the gravitational mass and the baryonic mass, one can
in principle use the observed mass and the small mass range expected for an
e-capture progenitor ($M_0 \sim 1.366-1.375M_\odot$) to constrain the
EOS\cite{pdl+05}. However, alternative ways of producing such light
neutron stars, e.g.~via ultra-stripped Type Ic Supernovae from close
binary evolution\cite{tlm+13}, have been proposed also.

The Double Pulsar may also allow us to actually measure the
moment-of-inertia of a neutron star. As this combines the mass and the
radius of a neutron star in one observable directly, such a measurement
would be very significant in determining the correct
EOS\cite{lbk+04,ls05}. Indeed, a measurement of the moment-of-inertia of
pulsar A in the Double Pulsar, even with moderate accuracy ($\sim
10$\%), would provide important
constraints.\cite{mbsp04,bbh05,ls05}  Recent timing results revealing
2PN-effects at the required level give hope that this goal can be
reached eventually\cite{ks+14}. See Ref.~\refcite{kw09} for a detailed
review on the prospects for making such measurement.

\section{Fast Radio Bursts, Revisited}

In Part I we introduced a new type of transient radio sources now
known as Fast Radio Bursts (FRBs). In the context of fundamental physics,
we are interested in exploring their nature on the one hand, and their
usage a probes on the other. As their origin is still unclear, we will
only attempt to give an overview of the existing, fast growing
literature. We start with looking at the origin of FRBs. 

All FRBs detected follow a perfect $\nu^{-2}$-dispersion law, as it is
expected from signal propagation in a cold ionized medium. In the
discussion, whether the signals are Galactic or extra-galactic,
Ref.~\refcite{lsm14} proposed FRBs may actually be Galactic flare
stars wherein the large dispersion measure is due to dense plasma in
low-mass star atmospheres, rather than a demonstration of a large
distance traversed.  However Refs.~\refcite{lg14,den14,kon+14} reject
this Galactic model using radiation transfer arguments; e.g. such high
plasma densities should produce enormous intrinsic absorption that
should render them undetectable, or produce free-free emission that is
not seen, or result in a break-down of the cold plasma dispersion law,
which contradicts observations. Moreover, a number of FRBs  also
show signs for interstellar scattering. Where it has been possible
to measure (e.g. Ref.~\refcite{tsb+13}; see Fig.~\ref{fig:frb}), the frequency dependence of the
scattering time follows a $\nu^{-4}$-law, as expected for propagation
in interstellar and intergalactic space. With the
dispersion measure (typically vastly) exceeding the contribution
expected from the Milky Way, an extra-galactic origin is the most
likely explanation, with distances corresponding to redshifts of the
order of $z\sim1$ as inferred from an estimate of the intergalactic
free electron content\cite{tsb+13}.

From the combination of temporal brevity and great luminosity
(inferred from their large distances), we then immediately infer that
the sources must embody a physically extreme environment, likely
involving very high gravitational or magnetic
fields. Possibilities being discussed include interacting
magnetospheres of coalescing neutron stars, coalescing white dwarfs,
evaporating black holes, supernovae, and super-giant pulses (see Refs.~\refcite{tsb+13,cw15} and
references therein). More exotic models propose signals from (bare) strange
stars\cite{mppp14}, white holes\cite{brv14}, or super-conducting cosmic
strings\cite{ycst2014}. FRB emission must almost certainly be from a
coherent process as the implied brightness temperature for a thermal
process is impossibly large given the small size implied by the short
durations; considering less exotic models, one would there expect that
FRBs originate from some sort of compact object -- white dwarf,
neutron star or black hole.  One possibility that appears particularly
appealing based on expected event rates is giant magnetar
flares\cite{kon+14,lyu14}.  

Whatever FRBs turn out to be, as extragalactic transient signals, they promise
to be very useful cosmological probes. For example, their dispersion
measure enables us to account for the ionized baryons between us and
the FRB sources and to measure the curvature of spacetime through
which the radiation propagates. A number of recent publications
discuss these possibilities, many of which are very well summarized in
Ref.~\refcite{mkg+15}.  Generally, they fall in three categories,
i.e.~FRBs as locators of the “missing” baryons in the low $(z \le2)$
redshift universe, high-redshift cosmic rulers which have the
potential to determine the equation-of-state parameter $w$ over a
large fraction of cosmic history, or potential probes of primordial
(intergalactic) magnetic fields and turbulence. See
Ref.~\refcite{mkg+15} for more details.

\section*{Summary \& Conclusions}

As we hope we have shown in this review, the field of neutron star
research, and in particular radio pulsars, is extremely active, and
addresses a very broad diversity of physical and astrophysical
questions.  These range from the structure and physics of dense
supra-nuclear matter, to the fate and evolution of massive stars, to
the nature of gravity and the origins of the Universe and the
structure therein.  We challenge our Solvay conference colleagues to
identify an astrophysical area more replete with results and impact!
The future for this domain of astrophysical research appears to be
growing only brighter, buoyed in particular by the development and
proliferation of multiple major new radio telescopes, including LOFAR,
MWA, ALMA, Meerkat, ASKAP, CHIME, FAST, and in the next decade, SKA.
Moreover, this science goes hand-in-hand with the blossoming field of
astrophysical transients, whether considering magnetar bursts as
possible FRB progenitors, or considering NS-NS mergers as aLIGO/VIRGO
sources.  We look forward to either participating in or hearing the
results reported at the next Solvay astrophysics meeting (which will
hopefully take place in fewer years than have passed since the last!)
by which time we predict there will have been major discoveries in
gravitational wave physics, in gravity in general, and in neutron-star
astrophysics.


\end{document}